       \documentstyle[]{l-aa}
       \input psfig.tex
         
	\newcommand{\Msun}{\mbox{${\rm M_{\odot}}$}}
	\newcommand{\Zsun}{\mbox{${\rm Z_{\odot}}$}}

        \def\M12{${\rm M_{L,12}} $}
        \def\Mg2{${\rm Mg_{2}} $}

\def\smallskip{\vskip 8pt}
\def\littleskip{\vskip 6pt}

\begin{document}

  \thesaurus{}

   \title{Diagnostic of chemical abundances and ages from the line strength
 indices $\rm H_{\beta}$, $\rm \langle Fe \rangle$, and $\rm Mg_2$: 
          what do they really imply ? }
 
   \author{ R. Tantalo$^1$, A. Bressan$^2$, C. Chiosi$^1$}

   \institute{
   $^1$ Department of Astronomy, Vicolo dell' Osservatorio 5, 35122 Padua, 
            Italy\\
   $^2$ Astronomical Observatory, Vicolo dell' Osservatorio 5, 35122 Padua, 
            Italy}

   \offprints{R. Tantalo }

   \date{Received: April 1997.  Accepted: }

    \maketitle

    \markboth{}{}

\begin{abstract}

The line strength indices $\rm H_{\beta}$, $\rm Mg_{2}$, and $\rm \langle Fe
\rangle$, together with their gradients and dependence on the galaxy
luminosity (mass), observed in elliptical galaxies are customarily considered
as reliable indicators of systematic differences in age and  abundances of
$\rm Mg$ and $\rm Fe$. Furthermore, as the gradient in $\rm Mg_{2}$ happens to
be often steeper than the gradient in $\rm \langle Fe \rangle$, an enhancement
of $\rm Mg$ ($\alpha$-elements in general) with respect to $\rm Fe$ toward the
center of these systems or going from dwarf to massive ellipticals is
inferred. In this paper, we address the question whether or not the indices
$\rm Mg_{2}$ and $\rm \langle Fe \rangle$ (and their gradients) are real
indicators of chemical abundances and  enhancement of these or other effects
must be taken into account. We show that this is not the case because the
above indices are severely affected by the {\it unknown} relative percentage
of stars as function of the metallicity. In order to cast the problem in a
fully self-consistent manner, first we provide basic calibrations for the
variations $\rm \delta H_{\beta}$, $\rm \delta Mg_2$, and $\rm \delta \langle
Fe \rangle$  as a function of the age $\rm \Delta \log(t)$ (in Gyr),
metallicity $\rm \Delta \log(Z/Z_{\odot})$,  and $\rm \Delta [Mg/Fe]$. Second
and limited to three elliptical galaxies of the Carollo \& Danziger (1994a)
catalog, we analyze the implications of the gradients in $\rm Mg_2$ and $\rm
\langle Fe \rangle$ observed across these systems. It is shown from a
quantitative point of view how the difference $\rm \delta Mg_2$ and $\rm \delta
\langle Fe \rangle$ between the local (radial distance r) and  central values
of each index would translate into difference  $\rm \Delta [Mg/Fe]$,  $\rm
\Delta \log(Z/Z_{\odot})$, and $\rm  \Delta \log(t)$. Finally, the above
calibration is used to explore the  variations from galaxy to galaxy of the
nuclear values of $\rm H_{\beta}$, $\rm Mg_2$, and $\rm \langle Fe \rangle$
limited to a sub-sample of the Gonzales (1993) catalog. The differences $\rm
\delta H_{\beta}$, $\rm \delta Mg_2$, and $\rm \delta \langle Fe \rangle$ are
converted to differences $\rm \Delta \log(t)$, $\rm \Delta \log(Z/Z_{\odot})$,
and $\rm \Delta [Mg/Fe]$. Various correlations among the age, metallicity, and
enhancement variations are explored. In particular, we thoroughly examine the
relationship $\rm \Delta \log(t)-M_V$, $\rm \Delta \log(Z/Z_{\odot})-M_V$, and
$\rm \Delta [Mg/Fe]-M_V$, and advance the suggestion that the duration of the
star forming period gets longer or the age of the last episode of stellar
activity gets closer to the present at decreasing galaxy mass. This result is
discussed in the context of current theories of galaxy formation and
evolution. i.e. merger and isolation.   In brief, we conclude that none of
these can explain the results of our analysis, and suggest that the kind of
time and space dependent IMF proposed by Padoan et al. (1997) and the
associated models of elliptical galaxies elaborated by Chiosi et al. (1997)
should be at work.


\end{abstract}

\section{Introduction}

The line strength indices $\rm H_{\beta}$, $\rm Mg_{2}$, and  $\rm \langle
Fe\rangle$ and their gradients are considered as good indicators of age and
metallicity and customarily used either to infer these important physical
quantities  in specific regions of galaxies (the nucleus for instance), or to
derive the age and composition (abundances of $\rm Mg$ and $\rm Fe$) gradients
across these systems. Elliptical galaxies are the most popular targets of
these studies. In addition to this, since the  gradients in  $\rm Mg_{2}$ and
$\rm \langle Fe\rangle$ (Carollo \& Danziger 1994a,b; Carollo et al. 1993)
observed in elliptical galaxies have different slopes arguments are given for
an enhancement of $\rm Mg$ ($\alpha$-elements in general) with respect to $\rm
Fe$ toward the center of these galaxies. Finally,  the  inferred degree of
enhancement seems to increase passing from dwarfs to massive ellipticals (see
Worthey et al. 1994, Faber et al. 1992, and Matteucci 1997 for  recent reviews
of all these subjects and exhaustive referencing).

That $\rm H_{\beta}$ correlates with the age is not a surprise, because this
index is known to be most sensitive to the light emitted by stars at the
turnoff. However, since the properties of these latter also depend on the
chemical composition, we expect $\rm H_{\beta}$ to be  affected by chemical
parameters as well in a way that must be ascertained a priori in a
self-consistent fashion.

The bottom line to infer from $\rm Mg_{2}$ and $\rm \langle Fe\rangle$  an
enhancement in $\alpha$-elements rests on the implicit notion that these two
indices strongly depend on the metallicity in general and on the abundances of
Mg and Fe in particular and very little on the age. If this is the case, as
$\rm Fe$ is mainly produced by Type Ia supernovae (accreting white dwarfs in
binary systems) and in smaller quantities by Type II supernovae, the iron
abundance $\rm [Fe/H]$ continuously increases  as the galaxy ages. In contrast
only  Type II supernovae  contribute to oxygen and $\alpha$-elements.
Recalling that the mean lifetime of a binary system (Type Ia progenitors) is
$\geq 1$ Gyr, the $\rm [\alpha/Fe]$ ratios are expected to decrease during the
galaxy evolution. This means that to obtain a galaxy (or region of it)
enhanced in $\alpha$-elements the time scale of star formation over there must
be shorter than about 1 Gyr. This is a very demanding constraint on models of
galaxy formation and evolution.

In this paper, we address the question whether or not the indices $\rm Mg_{2}$
and $\rm \langle Fe \rangle$ (and their gradients) are real indicators of
chemical abundances and abundance ratios or other subtle effects must be taken
into account. More specifically, using the spherical  models of elliptical
galaxies with gradients of mass density, star formation  rate, and chemical
abundances developed by Tantalo et al. (1997) for which the  chemical
abundances are  known as a function of time and radial distance, we check how 
gradients in Mg and Fe (and their ratio) translate into gradients in  $\rm
Mg_2$ and $\rm \langle Fe\rangle$, and check whether a gradient in $\rm Mg_2$
steeper than the gradient in $\rm \langle Fe \rangle$ implies an enhancement
of the Mg with respect to Fe toward the center of these galaxies. We
anticipate here that, while these models are indeed able to match many  key
properties of elliptical galaxies, including the gradients in broad band
colors (see below), they lead to contradictory results as far as the gradients
in line strength indices are concerned (section 2). To understand the physical
cause of this odd behaviour of the models, we check the calibration in use and
the response of $\rm Mg_2$ to chemistry (section 3). The reason of the
contradiction is found to reside in the heavy dependence of the indices in
question on the metallicity distribution function $\rm N(Z)$ of the stellar
populations in a galaxy. In addition to this, the observed gradients in the
indices $\rm Mg_2$ and $\rm \langle Fe \rangle$ in particular, do not
automatically correspond to gradients in the Mg and Fe abundances (section 3).
In order to tackle the whole problem in  a fully self-consistent way we derive
new basic relationships providing the variation of the line strength indices
$\rm H_{\beta}$, $\rm Mg_2$, and $\rm \langle Fe \rangle$ of single stellar
populations (from which eventually the galactic indices are built up) as a
function of age $t$, total metallicity $Z$, and [Mg/Fe] (section 4). The
analysis is made for a few test galaxy of the Carollo \& Danziger (1994a,b)
sample. Second, we apply the above diagnostic to study the observational
variations of nuclear values of $\rm H_{\beta}$, $\rm Mg_2$ and $\rm \langle Fe
\rangle$ from galaxy to galaxy of the Gonzales (1993) sample and translate
these variations into differences in age, metallicity, and enhancement of
$\alpha$-elements (section 5). A preliminary ranking of galaxies as a function
of their last episode of star formation is attempted (section 6), The results
of the analysis are discussed at the light of the present-day understanding of
the mechanism of  formation and evolution of elliptical galaxies: i.e.
isolation or merger (section 7), and the SN-driven galactic wind or variable
IMF schemes (section 8). Finally, some concluding remarks are drawn (section
9).

\section{The reference model }

\subsection{The basic assumptions}

Elliptical galaxies are described as spherically symmetric systems whose mass
density decreases outward. The density of luminous material as a function of
the radial distance and the corresponding gravitational potential are derived
from Young (1976). We assume that each shell contains about 5\% of the total
mass in the luminous material $\rm M_L$. The mass distribution and
gravitational potential of the dark-matter as a function of the radial
distance are derived from the density profile of Bertin et al. (1992) however
adapted to the Young formalism. The total mass $\rm M_D$ and radius $\rm R_D$
of the dark-matter component are assumed to be $\rm 5\times M_L$ and  $\rm
5\times R_L$, respectively.

In order to simulate the collapse of luminous material into the potential well
of dark-matter (whose mass is assumed to be constant in time) the infall scheme
is adopted and the density of luminous material (gas) is let grow with time at
the rate:

\begin{equation}
\frac{d{\rho_{L}}(r,t)}{dt} = \rho_{L0}(r) e^{-\frac{t}{\tau(r)}}
\label{drho}
\end{equation}

\noindent
where $\tau(r)$ is the time scale of gas accretion (in principle it can be a
function of the radial distance), and $\rho_{L0}(r)$ is fixed by imposing that
at the present-day age of the galaxy $T_{G}$ the density of luminous material
in each shell has grown to the value given by the Young profile.

A successful description of the gas accretion phase is possible adapting to
galaxies the radial velocity law describing the final collapse of the core in
a massive star, i.e. free-fall in all regions external to a certain value of
the radius ($v(r) \propto r^{-\frac{1}{2}}$) and homology inside ($v(r)
\propto r$). This picture is also confirmed by numerical  calculations of
dynamical models with the Tree-SPH technique (cf. Carraro et al. 1997). This
simple scheme allows us to derive the radial dependence of $\tau(r)$ as a
function of some arbitrary time scale. For this latter we adopt the mean
free-fall time scale of Arimoto \& Yoshii (1987).

\noindent
The rate of star formation (SFR) follows the standard Schmidt law

\begin{equation}
\Psi(r,t) = \nu(r) {\rho_{Lg}}(r,t)
\label{esfr}
\end{equation}

\noindent
where $\rho_{Lg}(r,t)$ is the local gas density and $\nu(r)$ is the specific
efficiency. For this latter, we have adopted the formulation by Arimoto \&
Yoshii (1987).

\begin{figure*}[]
\psfig{file=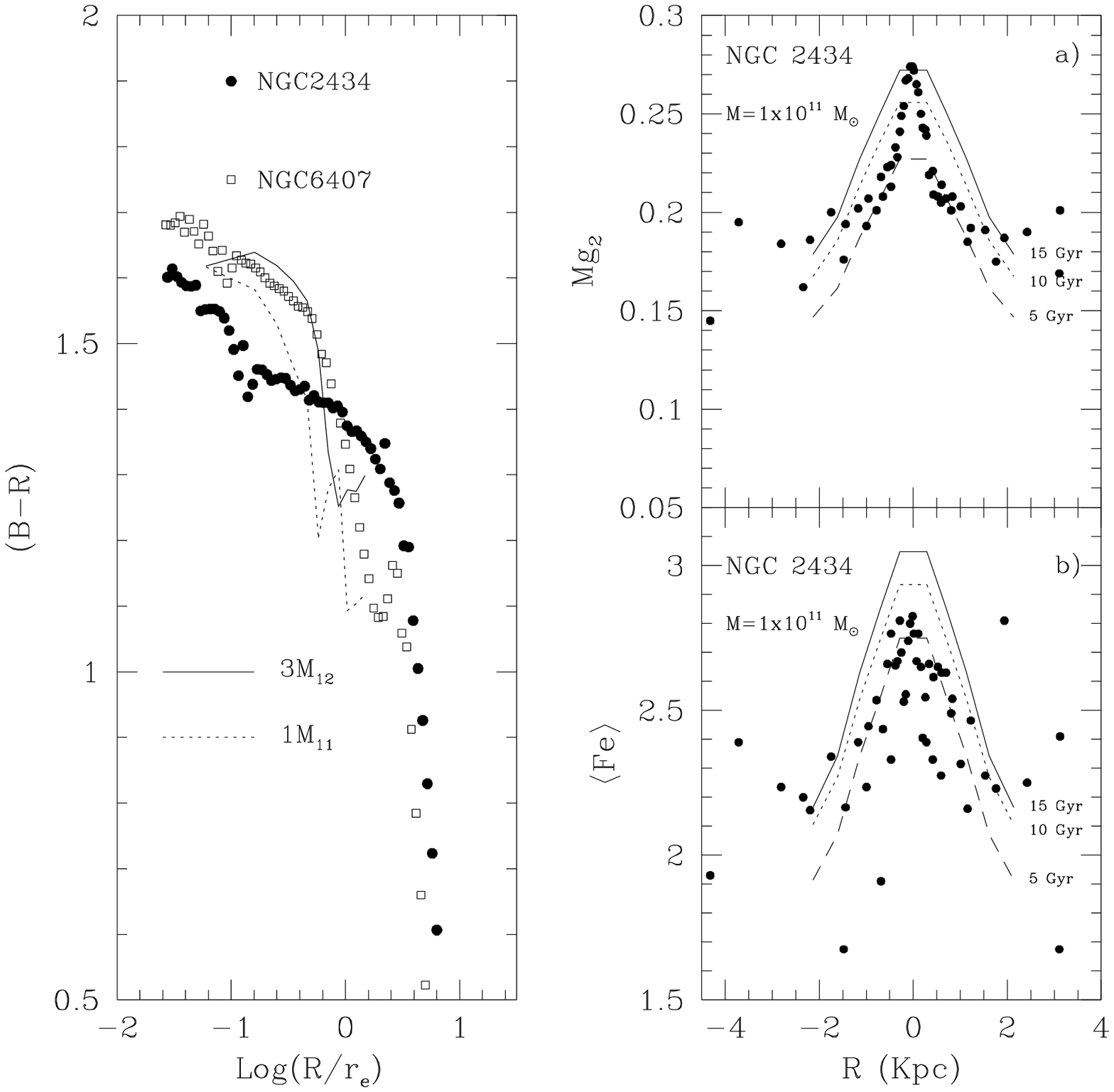,height=9.0truecm,width=17.0truecm}
\caption{Gradients in colors and line strength indices for the galaxies
NGC~2434 and NGC~6407 (Carollo \& Danziger 1994a), filled and open circles,
respectively. {\it Left Panel}: comparison with the theoretical gradients in
(B--R) for models of different mass and same age (15 Gyr).  {\it Right Panel}:
comparison with the theoretical gradients in line strength indices for the
$\rm 1 \times 10^{11} M_{\sun}$ model which has nearly the same $\rm M/L_{B}$
ratio as NGC~2434.}
\label{mod_car}
\end{figure*}

The chemical evolution of elemental species is governed by the same set of
equations as in Tantalo et al. (1996) however adapted to the density formalism
and improved as far as the ejecta and the contribution from Type Ia and Type
II supernovae are concerned according to the revision made by Portinari et al.
(1997) to whom we refer.

The models allow for galactic winds triggered by the energy deposit from Type
I and II supernova  explosions and stellar winds from massive stars. To
evaluate the amount of energy stored into the interstellar medium by supernova
explosions we adopt the same cooling law as in Gibson (1994, 1996).

Finally, the line strength indices $\rm Mg_{2}$ and $\rm \langle Fe \rangle$
have been calculated adopting the calibrations by Worthey (1992) and Worthey
et al. (1994) as a function of $\rm [Fe/H]$, $\rm T_{eff}$ and { gravity} of
the stars.

The basic data for the two innermost shells of a typical galaxy with total
luminous mass $\rm 3\times 10^{12} M_{\odot}$ are given in Table~1, whereas the
evolution of a few chemical abundances limited to the central region ($\rm
r=0.06 R_{e}$) is shown in Fig.~\ref{x_age}. The content of Table~1 is as
follows: $\rm M_{L,12}$ is the asymptotic luminous mass in units of $\rm
10^{12} M_{\odot}$; $\nu$ is the efficiency of the SFR; $\tau$ is the time
scale of gas accretion in Gyr; $\rm Z_{max}$ and $\rm \langle Z \rangle$ are
the maximum and mean metallicity, respectively; $\rm G(t)$ and $\rm S(t)$ are
the fractionary gas and alive star densities, respectively (both are
normalized to the asymptotic mass $\rm M_{L,12}$); $\rm N_{enh}$ is the
percentage of $\alpha$-enhanced stars present in the model (see below).

These galactic models are able to reproduce (i) the slope of colour-magnitude
relation by Bower et al. (1992); (ii) the UV excess as measured by the colour
(1550--V), (iii) the mass to blue luminosity ratio $\rm (M/L_{B})_{\odot}$ of
elliptical galaxies. For all other details see Tantalo et al. (1997).

\begin{table}
\begin{center}
\caption{Basic features of the reference model.}
\small
\begin{tabular} {l| c | c }
\hline
\hline
 & &  \\
\multicolumn{1}{l|}{parameter} &
\multicolumn{1}{c|}{$1^{st}$ Shell} &
\multicolumn{1}{c}{$2^{nd}$ shell} \\
 & &  \\
\hline
 & &  \\
 $\rm M_{L,12}$          & 0.146  & 0.150  \\
 $\nu$               & 7.1    & 50.0   \\
 $\tau$              & 0.74   & 0.29   \\
 $\rm t_{g\omega}$       & 5.12   & 0.79   \\
 $\rm Z_{max}$           & 0.0964 & 0.0439 \\
 $\rm \langle Z \rangle$ & 0.0365 & 0.0286 \\
 $\rm G(t)$              & 0.004  & 0.010  \\
 $\rm S(t)$              & 0.845  & 0.874  \\
 & & \\
\hline
 & & \\
 $\rm N_{enh}$           & 45.7\% & 53.9\%  \\
 & & \\
\hline
\hline
\end{tabular}
\end{center}
\label{tab1}
\end{table}

\subsection{The gradients}

In Fig.~\ref{mod_car} we compare the theoretical and observational gradients
in broad-band colors (left panel)  and  line strength indices (right panel).
The radial distance is expressed in units of the effective radius and in $Kpc$
respectively. The data are from Carollo \& Danziger (1994a). Remarkably, while
these models match the gradient in broad band colors which reflects the
gradient in metallicity (cf. Table~1), they somewhat fail as far as the
gradient in line strength indices $\rm Mg_{2}$ and $\rm \langle Fe \rangle$
are concerned. In brief, while the central value of $\rm Mg_2$  and at some
extent its gradient are matched (see the top left panel of
Fig.~\ref{mod_car}), this is not the case of the $\rm \langle Fe \rangle$
index (bottom left panel of Fig.~\ref{mod_car}). The theoretical gradients are
$\rm [\Delta ln Mg_2/ \Delta R] \simeq -0.13$ and $\rm[ \Delta ln \langle Fe
\rangle / \Delta R] \simeq  -0.11$ with R the galactocentric distance in Kpc.
The gradients are nearly identical.

\begin{figure}
\psfig{file=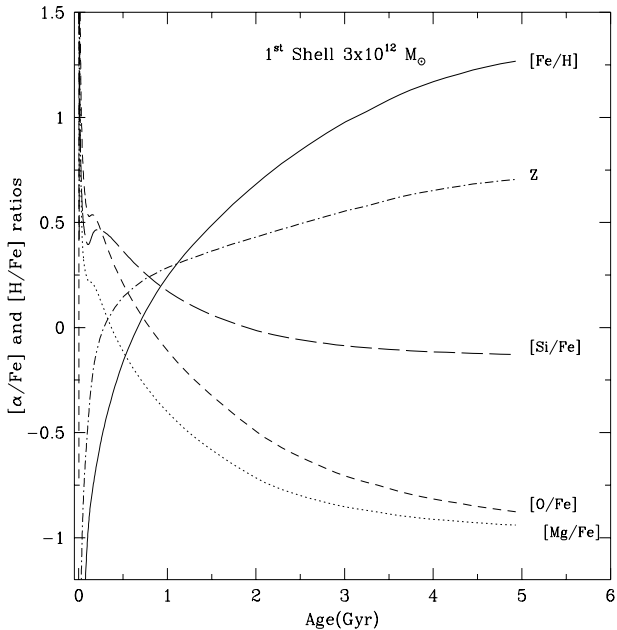,height=9.0truecm,width=8.5truecm}
\caption{Temporal evolution of four abundance ratios: $\rm [Fe/H]$ ({\em solid
line}), $\rm [Mg/Fe]$ ({\em dotted line}), $\rm [O/Fe]$ ({\em dashed line}),
and $\rm [Si/Fe]$ ({\em long-dashed line}). The {\em dot-dashed line} shows
the ratio $\rm Z/$\Zsun as a function of time. The ratios are in the standard 
notation}
\label{x_age}
\end{figure}

This is a surprising result because looking at the data of Table~1 the
duration of star formation was much longer in the central core than in the
outer shell (the same trend holds for all remaining shells not considered
here), which implies that the stars in the core are on the average less
$\alpha$-enhanced than in the more external regions (cf. the temporal
evolution of the elemental abundances shown in Fig.~\ref{x_age}), whereas the
gradients we have obtained seem to indicate a nearly constant ratio [Mg/Fe].

\begin{figure*}[]
\psfig{file=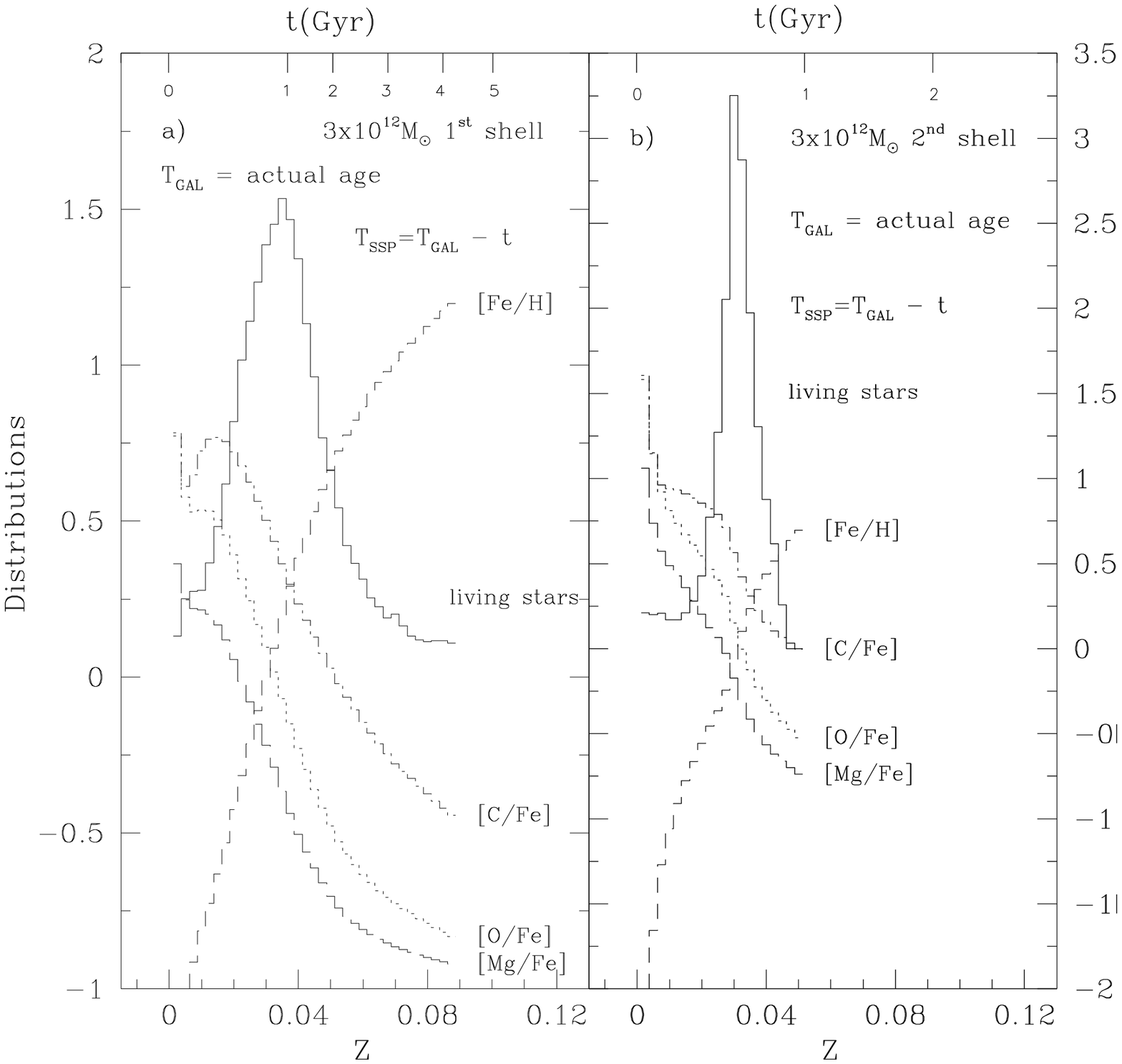,height=9.0truecm,width=17.0truecm}
\caption{{\it Panels (a)} and {\it (b)}: the number of living stars and
abundance ratio distribution per metallicity bin in the $1^{st}$ and the
$2^{nd}$ shells, respectively, of the galaxy with mass ($\rm 3\times 10^{12}
M_{\odot}$). The abundance ratios are in the standard notation.  The {\em
solid line} is distribution of living stars in units of $10^{11}\times
M_{\odot}$. The {\em dotted}, {\em dashed}, {\em long-dashed}, and {\em
dot-dashed} lines give the distribution per metallicity bin of $\rm [O/Fe]$,
$\rm [Fe/H]$, $\rm [Mg/Fe]$ and $\rm [C/Fe]$, respectively. The top scale
gives the birth-time $t=T_{G} - T_{SSP}$ in Gyr of a SSP of age $T_{SSP}$ in a
galaxy of age $T_{G}$.}
\label{x_nz}
\end{figure*}

In order to cast light on this contradiction we examined the variation of the
abundance ratios $\rm [Fe/H]$, $\rm [C/Fe]$, $\rm [O/Fe]$ and $\rm [Mg/Fe]$
(with respect to the solar value) as a function of the metallicity and time,
and the present-day number frequency distribution, $\rm N(Z)$ (thereinafter
the metallicity partition function),  of living stars in different metallicity
bins. The simultaneous inspection of the abundance ratios as a function of the
metallicity (and hence age) and their partition function $\rm N(Z)$ gives
immediately an idea of the fraction of stars still existing in the stellar mix
with metallicity above any particular value and with abundance ratios above or
below the solar value. These relationships are shown in the two panels of
Fig.~\ref{x_nz} (the left panel is for the central core, $1^{st}$ shell; the
right panel is for the $2^{nd}$  more external shell).

To further clarify the question we also look at the current age of the stellar
population stored in every metallicity (we remind the reader that the
metallicity in this model in a monotonic increasing function of the age, cf.
Fig.~\ref{x_age}). The top axis of Fig.~\ref{x_nz} shows the correspondence
between metallicity and birth-time of the stellar content of each metallicity
bin, shortly named single stellar population (SSP). The SSP birth-time is $t =
T_{G} - T_{SSP}$, where $T_{G}$ is the present-day galaxy age and $T_{SSP}$
is the current age of the SSP.
  
What we learn from this figure is that the external shell is truly richer in
$\alpha$-enhanced stars ($\sim 53.9\%$ of the total) than the inner one ($\sim
45.7\%$ of the total). These percentages $\rm N_{enh}$ are given in Table~1.
This confirms our expectation that these models should predict gradients in
line strength indices consistent with the gradients in abundances.

{\it Which is the reason for such unexpected contradiction? }

One may argue that the above disagreement results either from the particular
galactic model used to perform the comparison or the adoption of calibrations,
such as those by Worthey (1992) and Worthey et al. (1994), which include the
dependence on $\rm [Fe/H]$, $\rm T_{eff}$ and { gravity} but neglect the
effect of enhancing the  $\alpha$-elements. Therefore before proceeding
further we have to check whether the usage of  other calibrations for the line
strength indices and/or galactic models would change the above conclusion.

\begin{table}
\begin{center}
\caption{Basic properties of test models.  Model-A: late galactic wind and no 
enhancement of $\alpha$-elements.  Model-B: early galactic wind and 
enhancement of $\alpha$-elements. Model-C: recent burst of star formation, and
galactic wind,  and strong enhancement of $\alpha$-elements. }
\small
\begin{tabular}{l| c | c | c}
\hline
\hline
 & &  \\
\multicolumn{1}{l|}{parameter} &
\multicolumn{1}{c|}{Model A} &
\multicolumn{1}{c|}{Model B} &
\multicolumn{1}{c}{Model C} \\
 & & & \\
\hline
 & & & \\
 $\rm M_{L,12}$          & 0.146  & 0.146  & 0.146         \\
 $\nu$                   & 7.1    & 100.0  & 0.1 $\div$ 50 \\
 $\tau$                  & 0.74   & 0.05   & 0.10          \\
 $\rm t_{g\omega}$       & 5.12   & 0.39   & 3.58          \\
 $\rm Z_{max}$           & 0.0964 & 0.0713 & 0.0878        \\
 $\rm \langle Z \rangle$ & 0.0365 & 0.0279 & 0.0294        \\
 $\rm G(t)$              & 0.004  & 0.002  & 0.003         \\
 $\rm S(t)$              & 0.845  & 0.942  & 0.994         \\
 & & & \\
\hline
 & & & \\
 $\rm N_{enh}$           & 45.7\% & 85.2\% & 74.8\% \\
 & & & \\
\hline
\hline
\end{tabular}
\end{center}
\label{tab2}
\end{table}

\section{Testing the sensitivity of $\rm Mg_{2}$ to calibrations and chemistry}

To answer the question posed in the previous section, first we change the
calibration in usage adopting a new one in which the effects of [Mg/Fe] are
explicitly taken into account, and second we discuss different, {\it ad hoc}
designed, galactic models in which different levels of enhancement in
$\alpha$-elements are let occur by artificially changing the history of star
formation.

\subsection{Moving to calibrations containing {\rm [Mg/Fe]}}

Many studies have emphasized that line strength indices depend not only on the
stellar parameters $T_{eff}$ and gravity, but also on the chemical abundances
(Barbuy 1994, Idiart et al. 1995, Weiss et al. 1995, Borges et al. 1995). The
recent empirical calibration by Borges et al. (1995) include the effect of
different $\rm [Mg/Fe]$ ratios. The calibration is

\begin{displaymath}
\rm {\ln}{Mg_{2}} = -9.037 + 5.795\cdot\frac{5040}{T_{eff}} + 0.398\cdot \log{g}
+~~~~~~~~
\end{displaymath}
\begin{equation}
\rm ~~~~~~~~+0.389 \left[ \frac{Fe}{H} \right] - 0.16 \left[ \frac{Fe}{H}
\right]^{2} + 0.981 \left[ \frac{Mg}{Fe} \right] 
\label{mg2}
\end{equation}

\noindent
which holds for effective temperatures and gravities in the ranges $\rm 3800 K
< T_{eff} < 6500$ K and $\rm 0.7 < \log{g} < 4.5$. From this equation we are
able to analyse the effect of different $\rm [Mg/Fe]$ enhancement.

Using equation (\ref{mg2}) one has to adopt a different relation between the
global metallicity $\rm Z$ and [Fe/H] with respect to the standard one (cf.
Bertelli et al. 1994). In fact, if the ratio $[\alpha/\rm Fe]$ is enhanced
with respect to the solar value, at given total metallicity the net [Fe/H]
must be scaled accordingly. We find that the  general relation in presence of
$\alpha$-enhancement is

\begin{equation}
\rm 
\left[\frac{Fe}{H} \right] = \log{\left(\frac{Z}{Z_{\odot}}\right)} -
 \log{\left(\frac{X}{X_{\odot}}\right)} -
0.8\left[\frac{\alpha}{Fe}\right] - 0.05\left[\frac{\alpha}{Fe} \right]^{2}
\label{feh}
\end{equation}

The above relations for $\rm Mg_2$ and $\rm [Fe/H]$ are used to generate
new SSPs and galactic models in which not only the chemical abundances are
enhanced with respect to the solar value but the effect of this on the line
strength indices is taken into account in a self-consistent manner.

\subsection{Three artificial galaxies}

We present here three galactic models that in virtue of their particular 
history of star formation, have  different chemical structures and degree of
enhancement in $\alpha$-element.
\littleskip
  
{\it Model-A: late galactic wind}.
This  case  has late galactic wind ($\sim 5.12$ Gyr), which means that the
Type Ia dominate the enrichment in Fe of the gas, and the ratio [$\alpha/\rm
Fe$] is solar or below solar for most of the time. This model is actually the
central region of the $3\rm \times 10^{12} M_{\odot}$ galaxy presented above.
The percentage ($\rm N_{enh}$) of $\alpha$-enhanced stars that are still alive
at the age of 15 Gyr amounts to 45.7\%.
\littleskip

{\it Model-B: early galactic wind}.
In order to enhance the relative abundance of elements from Type II supernovae
we arbitrarily shortened the duration of the star forming period. To this aim,
in the central region of the same galaxy, the efficiency of star formation has
been increased $\nu=100$) and the infall time scale decreases ($\tau$=0.05
Gyr) so that the galactic wind occurs much earlier (at 0.39 Gyr) than in the
previous case. The material (gas and stars) of Model-B is therefore strongly
enhanced in $\alpha$-elements. The percentages ($\rm N_{enh}$) of
$\alpha$-enhanced stars that are still alive at the age of 15 Gyr amounts to 
87.2\%.
\littleskip

\begin{figure}
\psfig{file=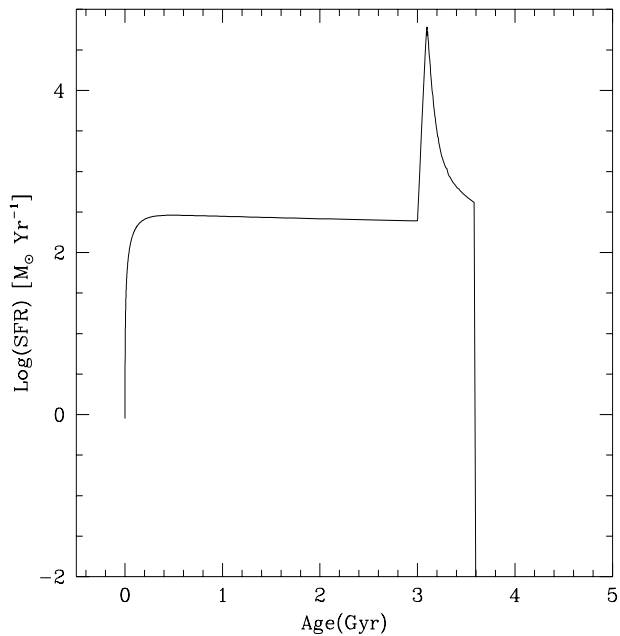,height=9.0truecm,width=8.5truecm}
\caption{{\it Model-C}: the star formation rate as a function of the time for
the central region of the galaxy model with $\rm 3\times 10^{12}\Msun $. At
the age of 3 Gyr the efficiency of the SFR is let increase  from $\nu = 0.1$
up to $\nu = 50$ over a time scale of  $10^{8}$yr. The parameters of this
model are given in Table~2.}
\label{sfr}
\end{figure}

\begin{figure}
\psfig{file=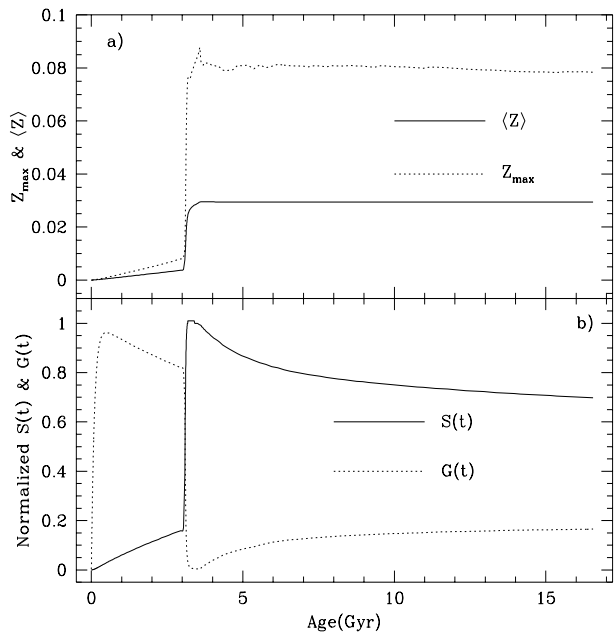,height=9.0truecm,width=8.5truecm}
\caption{{\it Model-C}: Panel (a) shows maximum and mean metallicity {\em
dotted} and {\em solid line}, respectively. Panel (b) shows the fractionary
density of gas $\rm G(t)$ and living stars $\rm S(t)$ as a function of time,
{\em dotted} and {\em solid line}, respectively.}
\label{chem}
\end{figure}

{\it Model-C: recent burst of star formation}. 
A third possibility is considered, in which a burst of star formation can
occur within a galaxy that already underwent significant star formation and
metal enrichment during its previous history. This model (always limited to
the central region of the galaxy) has nearly constant star formation rate from
the beginning, but at the age (arbitrarily chosen) of 3 Gyr it is supposed to
undergo a sudden increase perhaps induced by a merger. Accordingly, the
specific efficiency of star formation $\nu$ goes from $\nu=0.1$ to $\nu=50$
over a time scale of $10^{8}$~yr. The initial nearly constant stellar activity
is secured adopting a long time scale of gas accretion in the infall scheme
($\tau=10$ Gyr). The rate of star formation as a function of time in units of
\Msun$\rm yr^{-1}$ is given in Fig.~\ref{sfr}. Soon after the intense period
of star formation, the galactic wind occurs thus halting  star formation and
chemical enrichment. The basic chemical properties of the model as a function
of the age are shown in Fig.~\ref{chem}. This displays the maximum ({\em
dotted line}) and mean metallicity ({\em solid line}), and the fractionary
mass of gas $\rm G(t)$ ({\em dotted line}) and stars $\rm S(t)$ ({\em solid
line}) both normalized to the asymptotic galactic mass in units of
$10^{12}$\Msun. The percentage ($\rm N_{enh}$) of $\alpha$-enhanced stars that
are still alive at the age of 15 Gyr amounts to 74.5\%.

\littleskip

\begin{figure}
\psfig{file=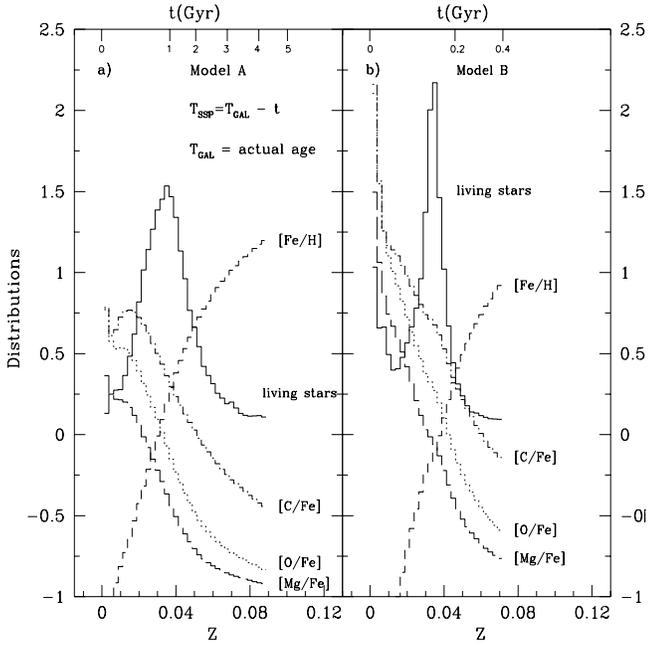,height=9.0truecm,width=8.5truecm}
\caption{The parttition function $\rm N(Z)$  and abundance ratios distribution
per metallicity bin, for the Model-A (panel a) and Model-B (panel b). The {\em
solid line} is $\rm N(Z)$ in units of $10^{11}$ at the gae of 15 Gyr. The {\em
dotted}, {\em dashed}, {\em long-dashed}, and {\em dot-dashed} lines give the
distribution per metallicity bin for $\rm [O/Fe]$, $\rm [Fe/H]$, $\rm [Mg/Fe]$
and $\rm [C/Fe]$, respectively. The abundance ratios are in the standard
notation. The top scale gives the birth-time $t=T_{G}-T_{SSP}$ of a SSP with
age $T_{SSP}$ in a galaxy with age $T_G$.}
\label{x_nz_ab}
\end{figure}

\subsection{The experiments} 

\begin{figure}
\psfig{file=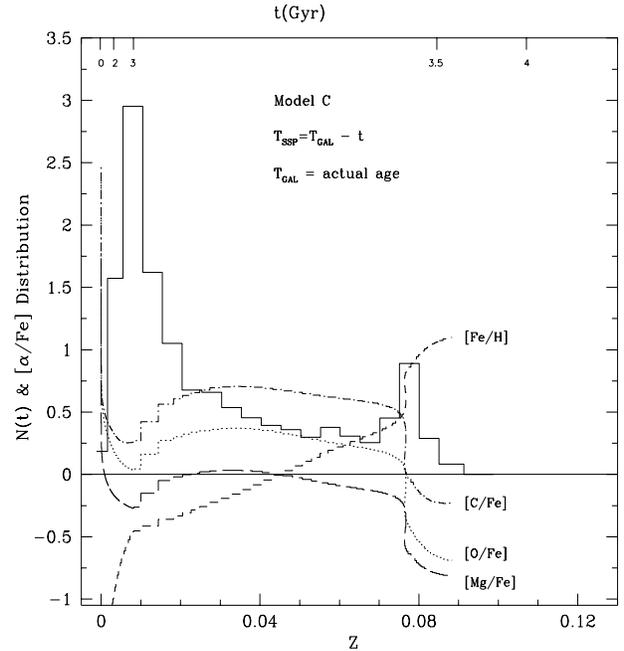,height=9.0truecm,width=8.5truecm}
\caption{The number of living stars and abundance ratios distribution per
metallicity bin, for the Model-C. The abundance ratios are in the standard
notation. The {\em solid line} is distribution of living stars in units of
$10^{11}$. The {\em dotted}, {\em dashed}, {\em long-dashed}, and {\em
dot-dashed} lines give the distribution per metallicity bin for $\rm [O/Fe]$,
$\rm [Fe/H]$, $\rm [Mg/Fe]$ and $\rm [C/Fe]$, respectively. The top scale
gives the birth-time $t=T_{G}-T_{SSP}$ of a SSP with age $T_{SSP}$ in a galaxy
with age $T_G$.}
\label{modc}
\end{figure}

The basic data for the three models in question are summarized in Table~2,
whereas their histories of chemical abundances together with the present-day
partition function $\rm N(Z)$ are  shown in Figs.~\ref{x_nz_ab} (left panel
for Model-A, (right panel for Model-B) and ~\ref{modc} for Model-C. These
figures are the analog of Fig.~\ref{x_nz}.

The distribution of chemical abundances (and their ratios) as a function of
the total metallicity and/or time  is used  to calculate SSPs of different
$\rm Z$, $\rm [Fe/H]$, $\rm [O/Fe]$, $\rm [Mg/Fe]$ for which with the aid of
equations (\ref{mg2}) and (\ref{feh}) we derive the corresponding $\rm Mg_2$
index. 

The procedure goes as follows: assigned the total metallicity $Z$ we read from
Figs.~\ref{x_nz_ab} or ~\ref{modc} the ratios $\rm [C/Fe]$, $\rm [O/Fe]$, and
$\rm [Mg/Fe]$. We derive $\rm [Fe/H]$ from equation (\ref{feh}) and insert
$\rm [O/Fe]$ and $\rm [Fe/H]$ into equation (\ref{mg2}). It goes without
saying that $\rm [Fe/H]$ read derived from Figs.~\ref{x_nz_ab} or ~\ref{modc}
and equation (\ref{feh}) are mutually consistent by definition.

Even if we will always refer to the ratio [Mg/Fe], to quantify the level of
enhancement in $\alpha$-elements we prefer to adopt the ratio $\rm [O/Fe]$,
because our chemical models of galaxies   somewhat underestimate the ratio
$\rm [Mg/Fe]$ as compared to the observational value. This is due to the fact
that the chemical models of galaxies by  Portinari et al. (1997)  over-estimate
the production of Fe by the Type II supernovae (cf. Thielemann et al. 1993,
1996). This marginal drawback of the chemical model does not however affect
the conclusion of this analysis.

The method to calculate the line strength index $\rm Mg_2$ is the same as in
Bressan et al. (1996) but for the different calibration.

Table~3 shows the values of $\rm [Fe/H]$ and $\rm [O/Fe]$ assigned to the SSPs
according to chemical structure of Model-A, Model-B and Model-C, and for
purposes of comparison to the reference SSPs with no enhancement at all.

\begin{figure}
\psfig{file=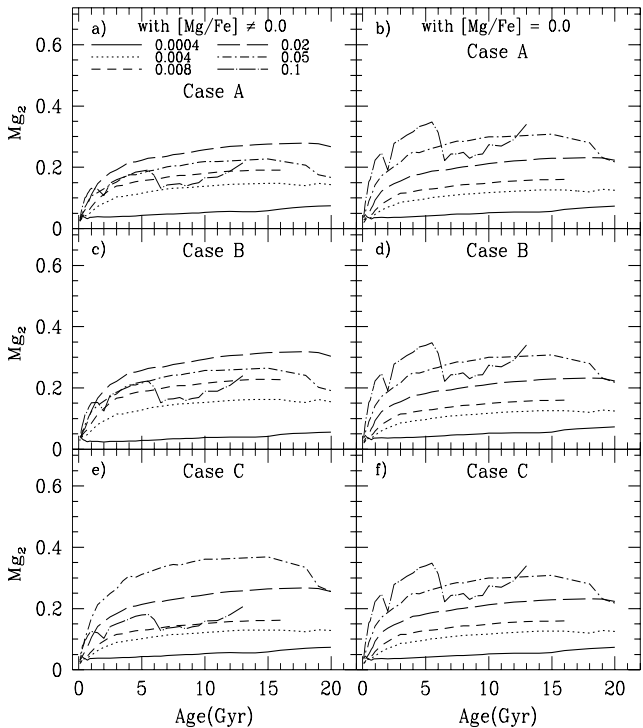,height=10.5truecm,width=9.0truecm}
\caption{{\it Panels (a)}, {\it (c)} and {\it e)} show the $\rm Mg_{2}$ index
evolution for SSPs with different metallicity (as indicated) and the
assumption of enhancement in $\alpha$-elements. {\it Panels (b)}, {\it (d)}
and {\it (f)} show the same but without enhancement of $\alpha$-elements.}
\label{mg2_ssp}
\end{figure}

The temporal evolution of $\rm Mg_{2}$ index for the SSPs with different total
metallicity and  ratios [O/Fe], and [Fe/H] given in Table~3, is shown in the
six panels of Fig.~\ref{mg2_ssp}.

\begin{table*}
\begin{center}
\caption{$\rm [O/Fe]$ and $\rm [Fe/H]$ ratios  for SSPs with enhancement of
$\alpha$-elements according to Model-A, Model-B and Model-C. The same ratios
for the reference SSPs with no enhancement all are also shown.}
\scriptsize
\begin{tabular*}{110mm} {l| c c| c c| c c| c c}
\hline
\hline
 & & & & & & & & \\
\multicolumn{1}{l|}{Z} &
\multicolumn{2}{c|}{Model-A} &
\multicolumn{2}{c|}{Model-B} &
\multicolumn{2}{c|}{Model-C} &
\multicolumn{2}{c}{Reference SSPs} \\
\hline
 & & & & & & & & \\
 & [O/Fe] & [Fe/H] & [O/Fe] & [Fe/H] & [O/Fe] & [Fe/H] & [O/Fe] & [Fe/H] \\
 & & & & & & & & \\
\hline
 & & & & & & & & \\
 0.0004 & +0.8   & --2.38 & +3.28  & --4.87 & +0.51   & --2.13 & 0.0 & --1.71 \\
 0.004  & +0.6   & --1.22 & +1.72  & --2.23 & +0.14   & --0.82 & 0.0 & --0.71 \\
 0.008  & +0.5   & --0.80 & +1.22  & --1.44 & +0.03   & --0.42 & 0.0 & --0.39 \\
 0.02   & +0.4   & --0.30 & +0.72  & --0.57 & +0.30   & --0.22 & 0.0 &  +0.03 \\
 0.05   & --0.50 & +0.88  & --0.25 & +0.69  & +0.31   & +0.24  & 0.0 &  +0.50 \\
 0.1    & --0.80 & +1.55  & --0.59 & +1.40  & --0.87  & +1.60  & 0.0 &  +0.95 \\
 & & & & & & & & \\
\hline
\hline
\end{tabular*}
\end{center}
\label{tab3}
\end{table*}

{\it The ${\rm Mg_2}$ index of SSPs}. 
It is soon evident that in absence of enhancement in $\alpha$-elements (right
panels of Fig.~\ref{mg2_ssp}), $\rm Mg_{2}$ monotonically increases with the
metallicity, but for the extreme SSP with Z=0.1 for which the trend is
reversed at ages older than 5 Gyr. When the enhancement of $\alpha$-elements
is included the trend $\rm Mg_2$-metallicity $\rm Z$ is reversed soon after
$\rm Z$ get larger than solar (left panels of Fig.~\ref{mg2_ssp}).

\begin{figure}
\psfig{file=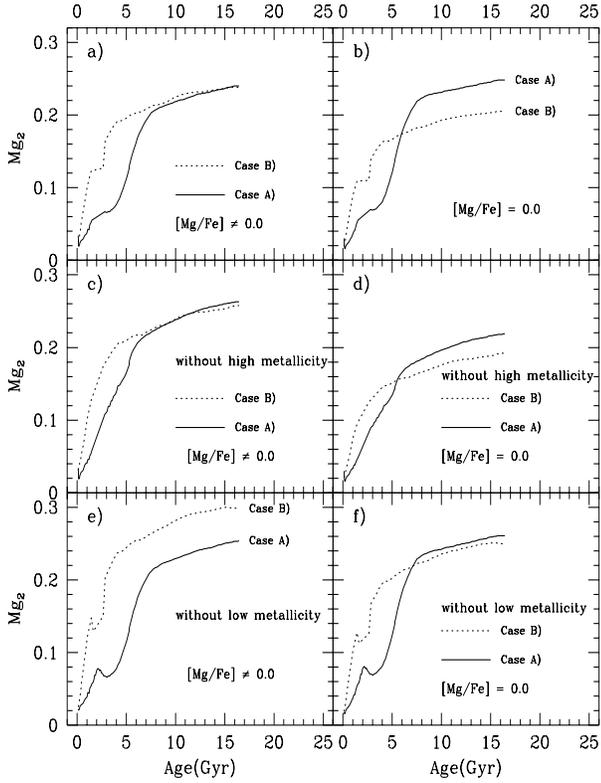,height=10.5truecm,width=9.0truecm}
\caption{ Evolution of the $\rm Mg_{2}$ index as a function of time. {\it
Panels (a)}, {\it (c)} and {\it (e)} show the evolution of the $\rm Mg_{2}$
index calculated including the effect of the chemical abundances, while {\it
Panels (b)}, {\it (d)} and {\it (f)} show the same but without the effect of
$\alpha$-enhancement. The {\em solid line} corresponds to Model-A and the
{\em dotted line} to Model-B.}
\label{mg2_ab}
\end{figure}

\begin{figure}
\psfig{file=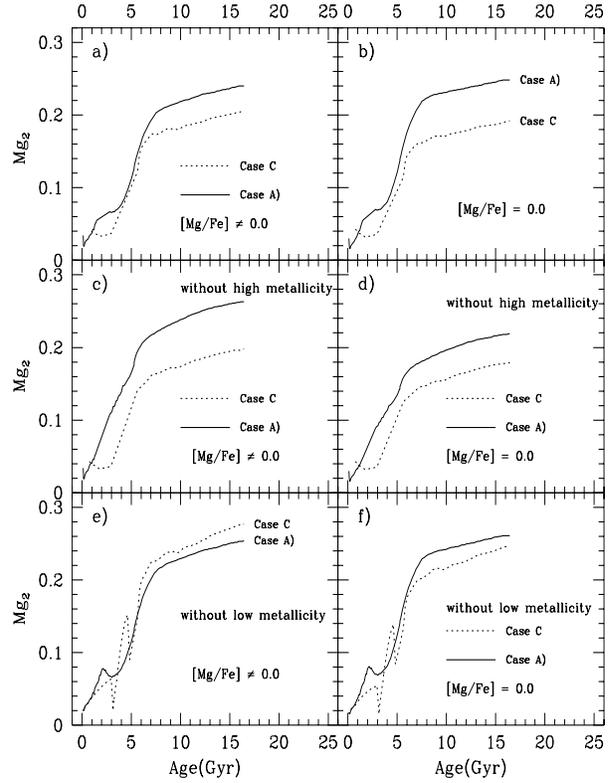,height=10.5truecm,width=9.0truecm}
\caption{Evolution of the $\rm Mg_{2}$ index as a function of time. {\it
Panels (a)}, {\it (c)} and {\it (e)} show the evolution of the $\rm Mg_{2}$
index calculated including the effect of the chemical abundances, while {\it
Panels (b)}, {\it (d)} and {\it (f)} show the same but without the effect of
$\alpha$-enhancement. The {\em solid line} corresponds to Model-A and the {\em
dotted line} to Model-C.}
\label{mg2_ac}
\end{figure}

{\it The ${\rm Mg_2}$ index of galaxies}. 
{\it What the results for the $\rm Mg_2$ index would be when applying these
SSPs with the Borges et al. (1995) calibration to model galaxies ? } The
situation is displayed in the various panels of Fig.~\ref{mg2_ab} for Model-A
and Model-B and in Fig.~\ref{mg2_ac} for Model-A and C. The combined analysis
of the chemical structures, partition functions $N(Z)$, and temporal
variations of the $Mg_2$ index of the model galaxies (top panels of
Figs.~\ref{mg2_ab} and ~\ref{mg2_ac}), the following remarks can be made

\begin{description}

\item [(i)] $\rm [Mg/Fe] \neq 0$: $\rm Mg_2$ in Model-A (late wind, no chemical
enhancement) is always weaker than in Model-B (early wind, significant
chemical enhancement). However, the difference is large for ages younger than
about 5 Gyr, and gets very small up to vanishing for older ages. $\rm Mg_2$ of
Model-C  is always weaker than Model-A and Model-B which means that the higher
(or comparable) enhancement in $\alpha$-elements of Model-C with respect to
the previous ones does not produce a stronger $\rm Mg_{2}$ index.

\item [(ii)] $\rm [Mg/Fe] = 0$: $\rm Mg_2$ in Model-A (late wind, no chemical
enhancement) is first weaker than in Model-B (early wind, significant chemical
enhancement) up to ages of about 5 Gyr, and then becomes significantly
stronger at older ages.  Finally, the $\rm Mg_{2}$ index of Model-C (burst of
star formation, and strong enhancement) is weaker or about equal to that of
Model-A and Model-B.
\end{description}

{\it The intensity of the $\rm Mg_2$ index does not simply correlate with the
abundance of $\rm Mg$ and in particular with the ratio {\rm [Mg/Fe]}. 
What causes this odd behaviour of the $ \rm Mg_2$ index as a function of the
age and underlying chemical structure of the model galaxy ? }
\littleskip

To answer this question, we have artificially removed from the partition
function $\rm N(Z)$ all the stars in certain metallicity bins and recalculated
the line strength indices for the three models.

Panels (c) and (d) of  Figs.~\ref{mg2_ab} and ~\ref{mg2_ac} show the results
when all stars with metallicity higher than Z=0.05 are removed. This is
motivated by the trend as function of the metallicity shown by the SSPs we
have already pointed out. The situation remains unchanged.

Likewise, panels (e) and (f) of Figs.~\ref{mg2_ab} and ~\ref{mg2_ac}  show the
same but when all stars with metallicity lower than Z=0.008 are removed. Now
the results change significantly. In the case of $\rm [Mg/Fe] \neq 0$, the $\rm
Mg_2$ index of Model-B is always much stronger than that of Model-A. In such a
case there is correspondence between the strength of the index and the amount
of enhancement in $\alpha$-elements. In contrast for $\rm [Mg/Fe] = 0$, at
younger ages $\rm Mg_2$ of Model-B is stronger at younger ages, but equals (or
even get weaker) that of Model-A at older ages. The same can be said comparing
Model-A to Model-C.

{\it What we learn from these experiments is that 
\begin{itemize}
\item{the index $\rm Mg_{2}$ heavily depends on the underlying partition
function $\rm N(Z)$, which of course is not known, so that inferring either
the abundance of Mg or the enhancement ratio [Mg/Fe] is a cumbersome affair.}

\item{The observational gradients in $\rm Mg_2$ (and $\rm \langle Fe \rangle$)
do not automatically imply gradients in chemical abundances or enhancement
ratios.}
\end{itemize}}

\section{Gradients across galaxies: a basic tool} 

{\it What do we learn from the gradients in line strength indices across 
galaxies?} 

Analyzing the gradients in $\rm H_{\beta}$ and $\rm [MgFe]$ measured in the
Gonzales (1993) sample of elliptical galaxies, Bressan et al. (1996)
translated those gradients into age and metallicity gradients and grouped the
galaxies of the sample in four sub-classes according to whether the nucleus
turned out to be older or younger and more or less metal-rich than the
external regions. The  majority of galaxies fell into the group with the
nucleus containing a fraction of stars younger and more metal-rich than the
external regions. In this section we intend to check whether or not the
gradients in age and metallicity suggested by Bressan et al. (1996) are
consistent with the gradients $\rm Mg_{2}$ and $\rm \langle Fe \rangle$
discussed in this paper.

To this purpose we calculate  the partial derivatives $\partial Mg_{2}$,
$\partial \langle Fe \rangle$ and $\partial H_{\beta}$ with respect to
metallicity, age, and $\rm [Mg/Fe]$ of SSPs

\begin{displaymath}
\frac{\partial Mg_{2}}{\partial \log{Z/Z_{\odot}}} \biggr|_{t,[\frac{Mg}{Fe}]}
                          \,\,\,\,\,\,\,\,\,\,\,
\frac{\partial Mg_{2}}{\partial [Mg/Fe]} \biggr|_{t,Z} \,\,\,\,\,\,\,\,\,\,\,
\frac{\partial Mg_{2}}{\partial \log{t}} \biggr|_{Z,[\frac{Mg}{Fe}]}
\label{dlgmgz}
\end{displaymath}
\noindent

\begin{displaymath}
\frac{\partial \langle Fe \rangle}{\partial \log{Z/Z_{\odot}}} \biggr|_{t,[\frac{Mg}{Fe}]}
                                    \,\,\,\,\,\,\,\,\,\,\,
\frac{\partial \langle Fe \rangle}{\partial [Mg/Fe]} \biggr|_{t,Z}
\,\,\,\,\,\,\,\,\,\,\,  \frac{\partial \langle Fe \rangle}{\partial \log{t}}
\biggr|_{Z,[\frac{Mg}{Fe}]}
\label{dlgfez}
\end{displaymath}

\begin{displaymath}
\frac{\partial H_{\beta}}{\partial \log{Z/Z_{\odot}}} \biggr|_{t,[\frac{Mg}{Fe}]}
                          \,\,\,\,\,\,\,\,\,\,\,
\frac{\partial H_{\beta}}{\partial [Mg/Fe]} \biggr|_{t,Z} \,\,\,\,\,\,\,\,\,\,\,
\frac{\partial H_{\beta}}{\partial \log{t}} \biggr|_{Z,[\frac{Mg}{Fe}]}
\label{dhbeta}
\end{displaymath}

The indices $\rm Mg_{2}$, $\rm \langle Fe \rangle$, and $\rm H_{\beta}$ of the
SSP are obtained as follows: given the total metallicity $Z$ and enhancement
ratio $\rm [Mg/Fe]$, we derive from eq. (\ref{feh}) the corresponding value of
$ [Fe/H]$ to be used in the calibration of Borges et al. (1995) for $\rm Mg_2$
(eq.~\ref{mg2}) and Worthey (1992) and Worthey et al. (1994) for $\rm \langle
Fe \rangle$ ($Fe_{5335}$ and $Fe_{5270}$) and $\rm H_{\beta}$. A summary of the
$\rm Mg_{2}$, $\rm \langle Fe \rangle$ and $\rm H_{\beta}$ indices of SSP with
different $Z$ and age is given in Table~4.

With the aid of these SSP we express the mean variation $\rm \delta Mg_{2}$,
$\rm \delta \langle Fe \rangle$ and $\rm \delta H_{\beta}$ as a function of
$\rm [Mg/Fe]$,  $\rm Z$,  and $\rm t$:

\begin{displaymath}
\rm \delta Mg_{2} = 0.1056 \times \Delta \left[ \frac{Mg}{Fe} \right] + 
\end{displaymath}
\begin{equation}
\rm ~~~~~~~~~ + 0.1765 \times \Delta \log{\left( \frac{Z}{\Zsun} \right)} 
           + 0.0890 \times \Delta \log{t}
\label{dmg}
\end{equation}

\begin{displaymath}
\rm \delta \langle Fe \rangle = -0.9606 \times 
                     \Delta \left[ \frac{Mg}{Fe} \right] + 
\end{displaymath}
\begin{equation}
\rm ~~~~~~~~~ + 1.8928 \times \Delta \log{\left( 
       \frac{Z}{\Zsun} \right)} + 0.7021 \times \Delta \log{t}
\label{dfe}
\end{equation}

\begin{displaymath}
\rm \delta H_{\beta} = -2079 \times \Delta \left[ \frac{Mg}{Fe} \right] + 
\end{displaymath}
\begin{equation}
\rm ~~~~~~~~~ - 0.3796 \times \Delta \log{\left( \frac{Z}{\Zsun} \right)} 
              - 1.2072 \times \Delta \log{t}
\label{dhb}
\end{equation}

\begin{table}
\begin{center}
\caption{New SSPs with different metallicity and $\rm [Mg/Fe]$.}
\scriptsize
\begin{tabular}{c| l| r| c| c| c| c}
\hline
\hline
 & & & & & & \\
\multicolumn{1}{c|}{Age(Gyr)} &
\multicolumn{1}{c|}{Z} &
\multicolumn{1}{c|}{$\rm [Mg/Fe]$} &
\multicolumn{1}{c|}{$\rm [Fe/H]$} &
\multicolumn{1}{c|}{$\rm Mg_{2}$} &
\multicolumn{1}{c|}{$\rm \langle Fe \rangle$} &
\multicolumn{1}{c}{$\rm H_{\beta}$} \\
\hline
 & & & & & & \\
15 & 0.004 &  +0.3 & --0.952 & 0.134 & 1.714 & 1.687 \\
15 & 0.004 &   0.0 & --0.707 & 0.125 & 1.931 & 1.710 \\
15 & 0.004 & --0.3 & --0.472 & 0.115 & 2.165 & 1.750 \\
15 & 0.02  &  +0.3 & --0.214 & 0.262 & 2.787 & 1.312 \\
15 & 0.02  &   0.0 &  +0.031 & 0.229 & 3.087 & 1.377 \\
15 & 0.02  & --0.3 &  +0.266 & 0.198 & 3.395 & 1.452 \\
15 & 0.05  &  +0.3 &  +0.253 & 0.365 & 3.606 & 1.197 \\
15 & 0.05  &   0.0 &  +0.497 & 0.308 & 3.937 & 1.279 \\
15 & 0.05  & --0.3 &  +0.733 & 0.258 & 4.276 & 1.368 \\
 & & & & & & \\
 5 & 0.004 &  +0.3 & --0.952 & 0.104 & 1.401 & 2.388 \\
 5 & 0.004 &   0.0 & --0.707 & 0.097 & 1.630 & 2.415 \\
 5 & 0.004 & --0.3 & --0.472 & 0.090 & 1.875 & 2.455 \\
 5 & 0.02  &  +0.3 & --0.214 & 0.211 & 2.431 & 1.884 \\
 5 & 0.02  &   0.0 &  +0.031 & 0.182 & 2.734 & 1.945 \\
 5 & 0.02  & --0.3 &  +0.266 & 0.155 & 3.050 & 2.018 \\
 5 & 0.05  &  +0.3 &  +0.253 & 0.308 & 3.235 & 1.648 \\
 5 & 0.05  &   0.0 &  +0.497 & 0.255 & 3.584 & 1.734 \\
 5 & 0.05  & --0.3 &  +0.733 & 0.210 & 3.943 & 1.828 \\
 & & & & & & \\
\hline
\hline
\end{tabular}
\end{center}
\label{tab4}
\end{table}

\begin{figure}
\psfig{file=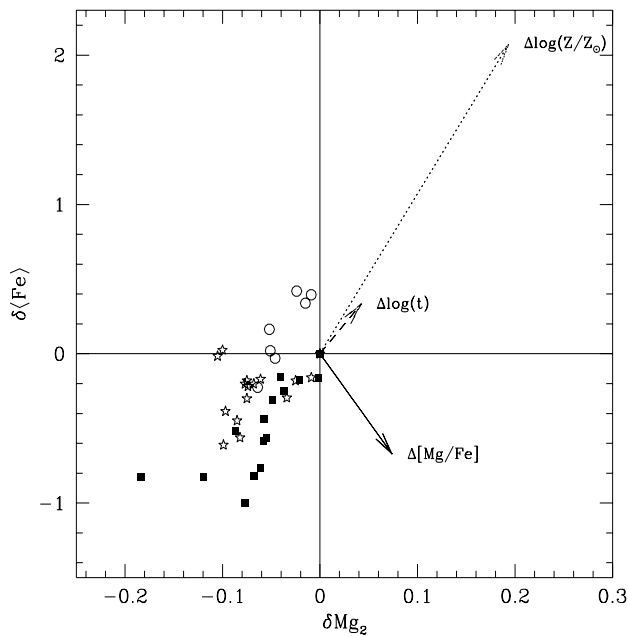,height=9.0truecm,width=8.5truecm}
\caption{The $\rm \delta \langle Fe \rangle$ versus $\rm \delta Mg_{2}$
relation. The three arrows centered on (0,0) indicate the {\it age,
metallicity, and enhancement vectors} as indicated. Along the age vector, the
age goes from 5 to 15 Gyr, along the metallicity vector $Z$ ranges from 0.004
to 0.05, and along the enhancement vector  $\rm [Mg/Fe]$ goes from $-0.3$ to
0.4 dex. The data are from Carollo \& Danziger (1994a) for NGC~2434, NGC~6407
and NGC~7192, {\em open circles}, {\em open stars}, and {\em filled squares},
respectively.}
\label{dlog}
\end{figure}

\noindent
To evaluate and visually show the size of $\rm \delta \langle Fe \rangle$ and
$\rm \delta Mg_{2}$ at varying $\rm [Mg/Fe]$, $Z$, and $t$, we calculate the
{\it age, metallicity,  and enhancement  vectors} shown in  Fig.~\ref{dlog}
for fixed variations of age,  metallicity, and $\rm [Mg/Fe]$: the age goes
from 5 to 15 Gyr, the metallicity from 0.004 to 0.05, and the $\rm [Mg/Fe]$
ratio from $-0.3$ to $0.4$  dex (the length of the vectors is $\rm \Delta
\log(Z) \sim 1.1$, $\rm \Delta  [Mg/Fe] \sim 0.7$ and $\rm \Delta \log(t) \sim
0.47$). The vectors are centered on (0,0), i.e. null variation.

There are two important  points to remark:
\begin{itemize}
\item The well known age-metallicity degeneracy given that the age and
metallicity vectors run very close each other.

\item The enhancement vector is almost orthogonal to the other two, which
allows us to separate its effects from the combined ones of age and
metallicity.
\end{itemize}

In the same diagram we show data for three galaxies, namely NGC~2434, NGC~6407
and NGC~7192 taken from Carollo \& Danziger (1994a). The displayed data are the
difference between the local value of each index (at any radial distance) and
its value at the center. Taking the galaxy NGC~7192 as a prototype, we first
derive an eye estimate of the mean slope of the data, second we translate  the
origin of the vectors to some arbitrary external point of the galaxy. The
result is that with respect to its external regions, the nucleus is either
more metal-rich and  older or more metal-rich and slightly younger,  and in
any case more enhanced in [Mg/Fe]. It goes without saying that if the nucleus
is older than the periphery a lower increase in the metallicity toward the
center is required than in the opposite alternative (younger nucleus). In such
a case the increase in metallicity must be large enough to compensate for the
opposite trend of the age. Similar considerations  apply also to the galaxies
NGC~2434 and NGC~6407.

{\it How this result conforms with the  Bressan et al. (1996)  analysis of
Gonzales (1993) galaxies in the $\rm H_{\beta}$ and $\rm [MgFe]$, suggesting
that in most galaxies the nucleus was younger (star formation lasted longer)
and more metal-rich than the external regions?}

\begin{table}
\begin{center}
\caption{Estimated gradients in age, metallicity and $\alpha$-enhancement
across two galaxies of the Carollo \& Danziger (1994a) sample. The nuclear
region is the one inside 5 arcsec from the centre. Ages are in Gyr.}
\scriptsize
\begin{tabular}{c| c c c c c c }
\hline
\hline
 & & & & & &  \\
\multicolumn{1}{c|}{NGC} &
\multicolumn{1}{c}{$\rm { {Mg_2}_N}$  } &
\multicolumn{1}{c}{$\rm { {\langle Fe \rangle}_N }$   } &
\multicolumn{1}{c}{$\rm { {H_{\beta}}_N  }$  } &
\multicolumn{1}{c}{$\rm { {Mg_2}_E}$    } &
\multicolumn{1}{c}{$\rm { {\langle Fe \rangle}_E }$   } &
\multicolumn{1}{c}{$\rm { {H_{\beta}}_E  }$  }\\
& & & & & &  \\ 
\hline
 & & & & & & \\
2434  & 0.235  &  2.65  &  2.2   &  0.194 &   2.42 &   1.8\\
3706  & 0.288  &  2.96  &  1.8   &  0.229 &   2.52 &   1.8\\
 & & & & & & \\
\hline
 & & & & & & \\
\multicolumn{1}{l|}{NGC} &
\multicolumn{1}{c}{$\rm \delta Mg_2$  } &
\multicolumn{1}{c}{$\rm \delta \langle Fe \rangle$ } &
\multicolumn{1}{c}{$\rm \delta H_{\beta}$ } &
\multicolumn{1}{c}{$\rm \Delta [{Mg\over Fe}]$  } &
\multicolumn{1}{c}{$\rm \Delta \log({Z\over Z_{\odot} }) $ } &
\multicolumn{1}{c}{$\rm \Delta \log(t) $}\\
& & & & & &  \\
\hline 
 & & & & & & \\
2434  & -0.041 &  -0.225 & -0.400 & -0.159 & -0.377 & 0.4776\\
3706  & -0.059 &  -0.440 &  0.000 & -0.106 & -0.332 & 0.1229\\
 & & & & & & \\
\hline
\hline
\end{tabular}
\end{center}
\normalsize
\label{tab5}
\end{table}

To answer the above question we make use of the $\rm H_{\beta}$ index, known
to be more sensitive to the age, and look at the parameter space $\rm
H_{\beta}$, $\rm Mg_2$ and $\rm \langle Fe \rangle$. Unfortunately, to our
knowledge this type of analysis is feasible only for two galaxies of the
Carollo \& Danziger (1994a,b) catalog for which all the data are available.
These are summarized in Table~5. The top part of the table shows the
observational data, i.e. the mean values of $\rm Mg_2$, $\rm \langle Fe
\rangle$, and $\rm H_{\beta}$ of the central region (indicated by a capitol N)
and the external region (labelled by a capitol E). The two regions are defined
by the visual inspection of the gradients in $\rm Mg_2$ and the transition
distance is seen to occur at 5 arcsec from the galactic centre (cf. Carollo \&
Danziger 1994a). The bottom part of Table~5 lists the difference $\rm \delta
Mg_2$, $\rm \delta \langle Fe \rangle$, and $\rm \delta H_{\beta}$ between 
the external and central values of the three indices, and the solution found
solving the system of equations (\ref{dmg}), (\ref{dfe}) and (\ref{dhb}). The
solution consists in the differences $\rm \Delta [Mg/Fe]$, $\rm \Delta
\log(Z/Z_{\odot})$, and $\rm \Delta \log(t)$ between the periphery and centre.
It turns out that both galaxies have the nuclear region containing stars more
enhanced in $\alpha$-elements, more metal-rich, and younger than the external
regions.

\begin{table*}
\begin{center}
\caption{ Basic data for galaxies of the Gonzales (1993) sample. The mean 
values of $\rm H_{\beta}$, $\rm Mg_2$ and $\rm \langle Fe \rangle$ are  
$\rm \overline{H_{\beta}}=1.72$,  $\rm \overline{Mg_{2}}=0.31$, and 
$\rm \overline{\langle Fe \rangle}=3.12$. }
\scriptsize
\begin{tabular*}{170mm}{l| c c c c c c r r r r r r c }
\hline
\hline
 & & & & & & & & & & & & & \\
\multicolumn{1}{l|}{NGC} &
\multicolumn{1}{c}{$\rm H_{\beta}$} &
\multicolumn{1}{c}{$\rm [MgFe]$} &
\multicolumn{1}{c}{$\rm Mg_{2}$} &
\multicolumn{1}{c}{$\rm Fe_{52}$} &
\multicolumn{1}{c}{$\rm Fe_{53}$} &
\multicolumn{1}{c}{$\rm \sigma_{0}$} &
\multicolumn{1}{c}{$\rm \delta Mg_2$} &
\multicolumn{1}{c}{$\rm \delta H_{\beta}$ } &
\multicolumn{1}{c}{$\rm \delta \langle Fe \rangle  $} &
\multicolumn{1}{c}{$\rm \Delta [{Mg \over Fe}]$} &
\multicolumn{1}{c}{$\rm \Delta \log({Z\over Z_{\odot} }) $} &
\multicolumn{1}{c}{$\rm \Delta \log(t)$ } &
\multicolumn{1}{c}{$\rm M_V $ }\\
 & & & & & & & & & & & & &\\
\hline
 & & & & & & & & & & & & &\\
 221 & 2.31 & 2.85 & 0.227 & 3.02 & 2.70 & 1.86 & -0.0884 &  0.5938 & -0.2603 & -0.2801 & -0.1303 & -0.4027& -16.64\\
 224 & 1.67 & 3.87 & 0.345 & 3.34 & 3.19 & 2.19 &  0.0296 & -0.0462 &  0.1447 &  0.0844 &  0.1250 & -0.0156&       \\ 
 315 & 1.74 & 3.74 & 0.323 & 3.08 & 3.14 & 2.51 &  0.0076 &  0.0238 & -0.0103 &  0.0485 &  0.0334 & -0.0386& -24.61\\ 
 547 & 1.58 & 3.76 & 0.328 & 3.19 & 3.50 & 2.37 &  0.0126 & -0.1362 &  0.2197 & -0.0533 &  0.0496 &  0.1064&       \\  
 584 & 2.08 & 3.54 & 0.295 & 3.21 & 2.96 & 2.29 & -0.0204 &  0.3638 & -0.0353 & -0.0496 &  0.0733 & -0.3159& -22.66\\ 
 636 & 1.89 & 3.57 & 0.285 & 3.32 & 3.04 & 2.20 & -0.0304 &  0.1738 &  0.0596 & -0.1711 & -0.0145 & -0.1099& -21.57\\ 
 821 & 1.66 & 3.65 & 0.333 & 3.29 & 3.03 & 2.28 &  0.0176 & -0.0562 &  0.0396 &  0.0688 &  0.0487 &  0.0194& -22.59\\ 
1700 & 2.11 & 3.53 & 0.296 & 3.31 & 3.06 & 2.36 & -0.0194 &  0.3938 &  0.0646 & -0.0884 &  0.1185 & -0.3483& -23.20\\ 
2300 & 1.68 & 3.84 & 0.352 & 3.27 & 3.04 & 2.40 &  0.0366 & -0.0362 &  0.0346 &  0.1756 &  0.1217 & -0.0385& -22.60\\ 
3377 & 2.09 & 3.23 & 0.283 & 2.91 & 2.66 & 2.03 & -0.0324 &  0.3738 & -0.3353 &  0.0302 & -0.0510 & -0.2988& -20.39\\  
 & & & & & & & & & & & & &\\
3379 & 1.62 & 3.69 & 0.324 & 3.22 & 3.00 & 2.31 &  0.0086 & -0.0962 & -0.0103 &  0.0396 & -0.0140 &  0.0773& -21.13\\ 
4374~V & 1.51 & 3.67 & 0.317 & 3.18 & 2.98 & 2.45 &  0.0016 & -0.2062 & -0.0403 &  0.0033 & -0.0937 &  0.1997& -22.73\\
4478~V & 1.84 & 3.56 & 0.243 & 4.04 & 2.89 & 2.11 & -0.0724 &  0.1238 & -0.1553 & -0.2999 & -0.2438 &  0.0258& -20.33\\ 
4552~V & 1.47 & 3.92 & 0.355 & 3.22 & 3.28 & 2.40 &  0.0396 & -0.2462 &  0.1297 &  0.1215 &  0.0705 &  0.1608& -21.68\\
4649 & 1.40 & 4.01 & 0.370 & 3.31 & 3.48 & 2.49 &  0.0546 & -0.3162 &  0.2747 &  0.1246 &  0.1349 &  0.1981& -22.82\\
4697~V & 1.75 & 3.36 & 0.300 & 3.31 & 3.05 & 2.21 & -0.0154 &  0.0338 &  0.0597 & -0.1071 & -0.0219 & -0.0027& -22.46\\ 
5638 & 1.65 & 3.63 & 0.325 & 3.17 & 2.85 & 2.19 &  0.0096 & -0.0662 & -0.1103 &  0.0962 & -0.0268 &  0.0467&       \\
5812 & 1.70 & 3.83 & 0.320 & 3.25 & 3.28 & 2.30 &  0.0046 & -0.0162 &  0.1447 & -0.0463 &  0.0510 &  0.0054&       \\
5813 & 1.42 & 3.53 & 0.311 & 3.08 & 2.83 & 2.31 & -0.0044 & -0.2962 & -0.1653 &  0.0200 & -0.1889 &  0.3013&       \\
5831 & 2.00 & 3.66 & 0.302 & 3.37 & 3.16 & 2.20 & -0.0134 &  0.2838 &  0.1447 & -0.1073 &  0.1158 & -0.2530& -21.85\\
 & & & & & & & & & & & & &\\
5846 & 1.45 & 3.76 & 0.340 & 3.18 & 2.99 & 2.35 &  0.0246 & -0.2662 & -0.0353 &  0.1173 & -0.0379 &  0.2122& -22.84\\ 
6127 & 1.50 & 3.76 & 0.329 & 3.07 & 3.03 & 2.38 &  0.0136 & -0.2162 & -0.0703 &  0.0809 & -0.0650 &  0.1856&       \\ 
6703 & 1.88 & 3.55 & 0.304 & 3.23 & 3.01 & 2.26 & -0.0114 &  0.1638 & -0.0003 & -0.0417 &  0.0298 & -0.1379&       \\
7052 & 1.48 & 3.77 & 0.338 & 3.18 & 3.18 & 2.44 &  0.0226 & -0.2362 &  0.0596 &  0.0648 & -0.0045 &  0.1859&       \\  
7454 & 2.15 & 2.84 & 0.236 & 2.72 & 2.44 & 2.03 & -0.0794 &  0.4338 & -0.5403 & -0.1172 & -0.2481 & -0.2611&       \\
7562 & 1.69 & 3.61 & 0.306 & 3.25 & 2.95 & 2.39 & -0.0094 & -0.0262 & -0.0203 & -0.0439 & -0.0497 &  0.0449&       \\ 
7619 & 1.36 & 3.93 & 0.370 & 3.36 & 3.48 & 2.48 &  0.0546 & -0.3562 &  0.2997 &  0.1079 &  0.1252 &  0.2371& -23.36\\ 
7626 & 1.46 & 3.78 & 0.366 & 3.15 & 3.27 & 2.40 &  0.0506 & -0.2562 &  0.0896 &  0.1984 &  0.0929 &  0.1489& -23.36\\ 
7785 & 1.63 & 3.66 & 0.324 & 3.14 & 3.15 & 2.38 &  0.0086 & -0.0862 &  0.0246 &  0.0241 &  0.0004 &  0.0671&       \\
 & & & & & & & & & & & & &\\
\hline
\hline
\end{tabular*}
\end{center}
\normalsize
\label{tab6}
\end{table*}

\section{Going from galaxy to galaxy}

{\it Given that central values of  $\rm H_{\beta}$, $\rm Mg_2$,  and $\rm
\langle Fe \rangle$ are know to vary from galaxy to galaxy, what the
implications are as far as the differences in age, metallicity, and
enhancement in $\alpha$-elements are concerned?}

To answer this question we make use of a small  sample of galaxies (29 objects
in total) selected by Worthey (1992) from the Gonzales (1993) catalog. These
galaxies are thought to possess the best determined line strength indices. The
data are presented in Table~6. Column (1) identifies the galaxy: those objects
whose name is followed by a V belong to the Virgo cluster. Columns (2) to (6)
give $\rm H_{\beta}$, $\rm [MgFe]$, $\rm Mg_{2}$, $\rm Fe_{5270}$ and $\rm
Fe_{5335}$ and velocity dispersion $\rm \sigma_{0}$ ($\rm Km~sec^{-1}$) for
the Re/8 data (nucleus) of the Gonzales (1993) galaxies. Columns  (7), (8) and
(9) show the differences $\rm \delta H_{\beta}$, $\rm \delta Mg_{2}$, and $\rm
\delta \langle Fe \rangle$ of the galaxy indices with respect to the mean
values (see below), respectively. Columns (10), (11), and (12) give the
difference with respect to the mean of the enrichment factor $\rm \Delta
[Mg/Fe]$, metallicity $\rm \Delta \log(Z/Z_{\odot})$, and age $\rm \Delta
\log(t)$ found for each galaxy. Finally, column (13) lists the total absolute
visual magnitude $M_V$.

For each set of data, we have calculated the mean values of $\rm Mg_{2}$, $\rm
\langle Fe \rangle$ and $\rm H_{\beta}$ to be used as the origin of a new
system of coordinates in which the differences $\rm \delta Mg_{2}$, $\rm
\delta \langle Fe \rangle$ and $\rm \delta H_{\beta}$ passing from galaxy to
galaxy are plotted and compared to the age, metallicity, and enhancement
vectors. The mean values are $\rm \overline{H_{\beta}}=1.72$, $\rm
\overline{Mg_{2}}=0.31$, and  $\rm \overline{\langle Fe \rangle}=3.12$.

\begin{figure}
\psfig{file=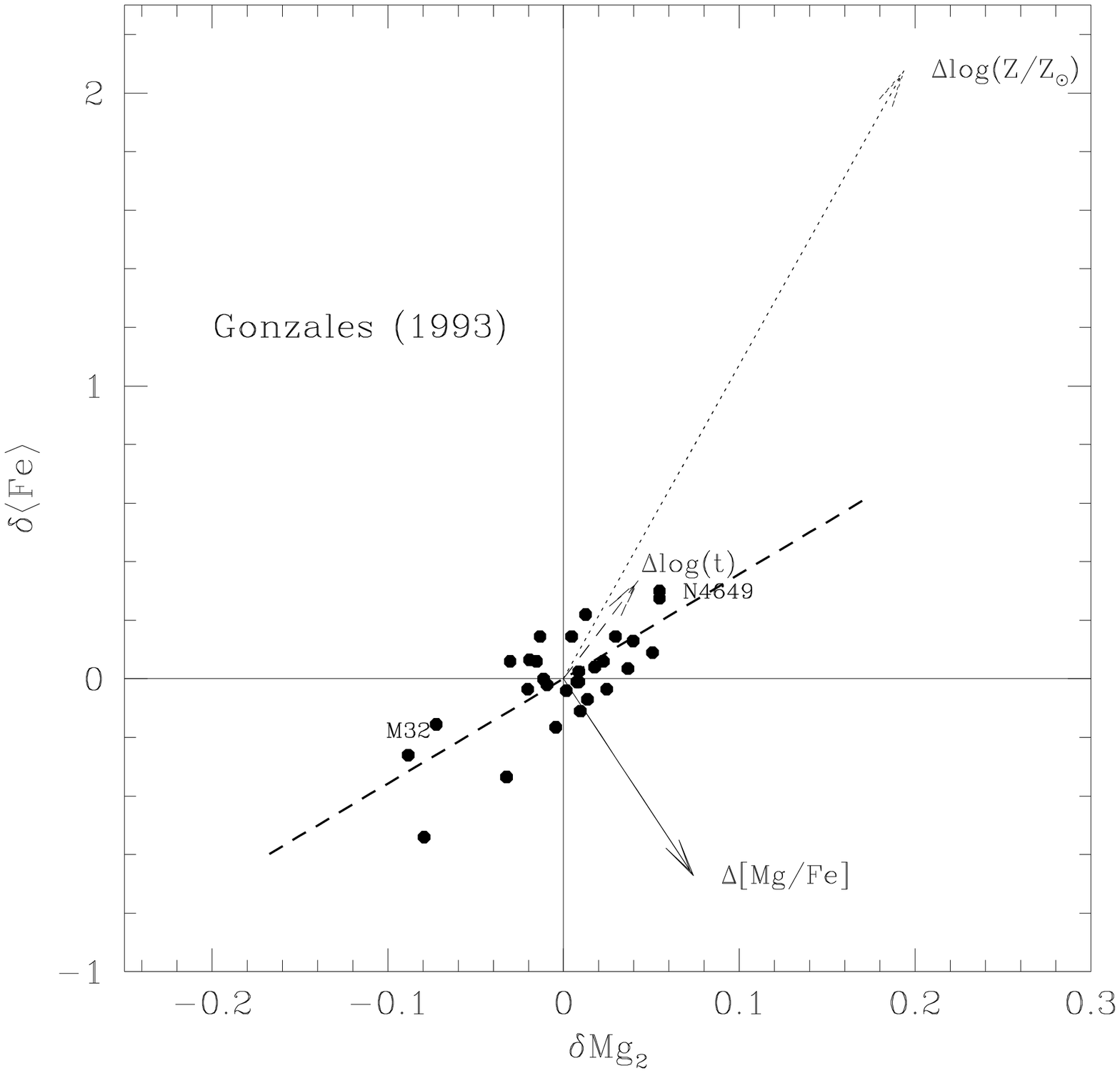,height=9.0truecm,width=8.5truecm}
\caption{The $\rm \delta \langle Fe \rangle$ versus $\rm \delta Mg_{2}$
relation. The three arrows centered on (0,0) are the {\it age, metallicity,
and enhancement vectors} as indicated. They have been calculated as in Fig.11.
The displayed data  are for each galaxy the difference between its central
value and the mean value of the sample. The position of M32 and  NGC~4649 are
indicated. The {\em thick dashed line} is the linear regression of the
data.}
\label{dgal_nuc}
\end{figure}

\begin{figure}
\psfig{file=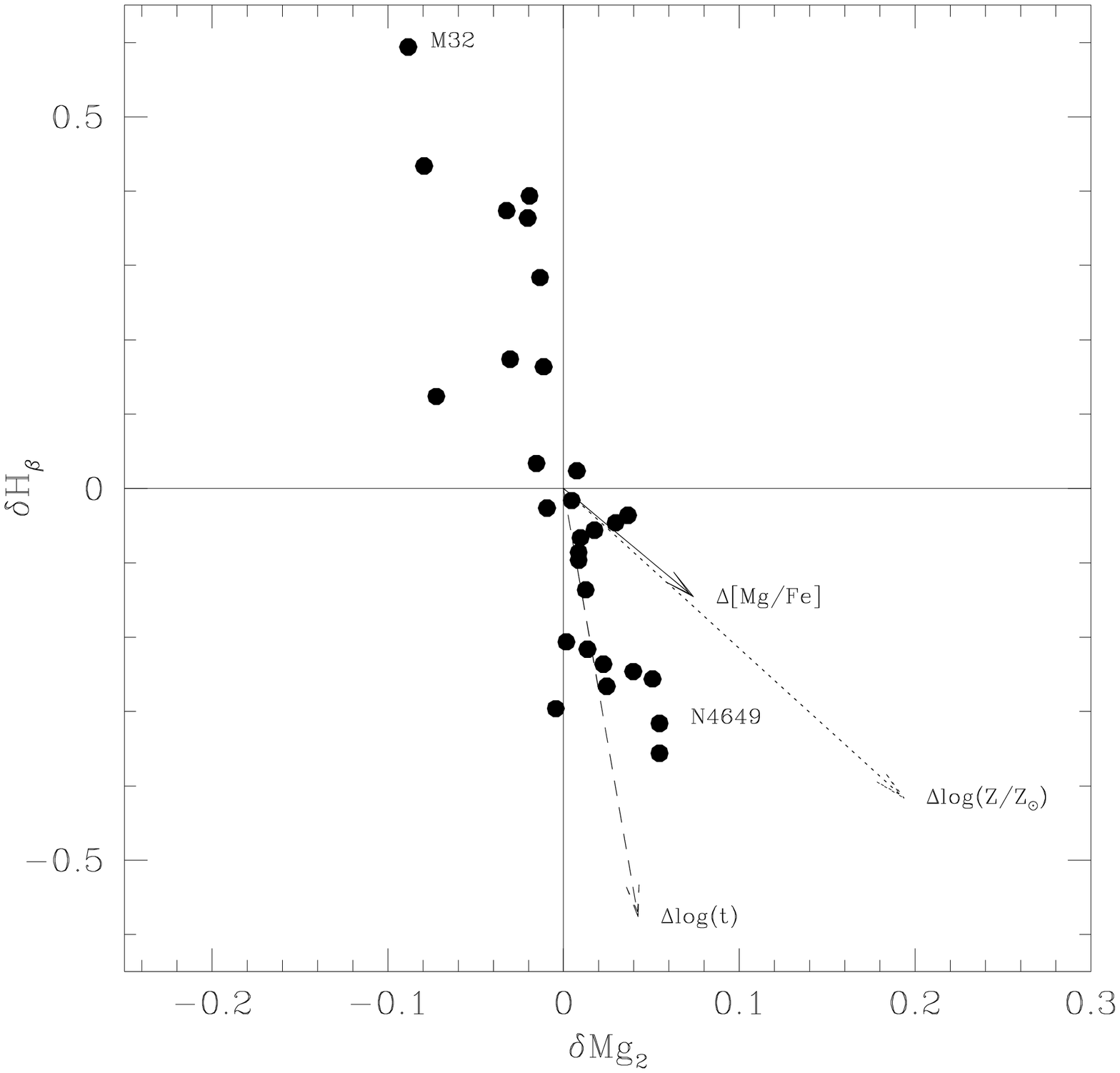,height=9.0truecm,width=8.5truecm}
\caption{The $\rm \delta H_{\beta}$ versus $\rm \delta Mg_{2}$ relation. The
three arrows centered on (0,0) are the {\it age, metallicity  and enhancement
vectors} as indicated. They have been calculated as in Fig.11. The data are
from Gonzales (1993) and for each galaxy the difference between its central
values and the mean value of the sample are displayed. The positions of M32
and NGC~4649 are indicated.}
\label{dgal_nucHb}
\end{figure}

\subsection{ The $\rm \delta Mg_{2}$-$\rm \delta \langle Fe \rangle$ and   
 $\rm \delta Mg_{2}$-$\rm \delta H_{\beta}$ planes}

The observational data for the differences $\rm \delta  Mg_{2}$ and $\rm
\delta \langle Fe \rangle$ are shown in Fig.~\ref{dgal_nuc}, in which we have
drawn the metallicity, enhancement, and age  vectors, whose length is the same
as in the previous analysis, and finally indicated the positions of M32 and
NGC~4649, the prototype galaxies in the discussion below. The thick
dashed line is the linear regression of the data.

Once again, age and metallicity cannot be safely separated, whereas this is
feasible for  $\rm [Mg/Fe]$.

In order to cope with the age-metallicity degeneracy encountered in the  $\rm
\delta Mg_{2}$ - $\rm \delta \langle Fe \rangle$ plane, we look at the $\rm
\delta Mg_{2}$ - $\rm \delta H_{\beta}$ plane because $\rm H_{\beta}$ is known
to be more sensitive to the age of the underlying stellar populations. This
is shown in Fig.~\ref{dgal_nucHb} where the moduli of the three vectors are:
$\rm \Delta \log(Z) \sim 1.1$, $\Delta \rm [Mg/Fe] \sim 0.7$ and  $\rm \Delta
\log(t) \sim 0.47$ In this figure the effect of $\rm [Mg/Fe]$ and metallicity
can not be isolated because the two vectors are coincident (degeneracy).

With the aid of the equations (\ref{dmg}) (\ref{dfe}) and (\ref{dhb}) and the
differences $\rm \delta H_{\beta}$, $\rm \delta Mg_2$, and $\rm \delta
\langle Fe \rangle$, we derive for each galaxy the variations in age $\rm
\Delta \log(t)$, metallicity $\rm \Delta \log(Z/Z_{\odot})$,  and enhancement
$\rm \Delta [Mg/Fe]$ with respect to the mean values. The results are listed
in columns (10), (11) and (12) of Table~6.

Looking at M32 and NGC~4649 as an example, we get the following results: for
M32 $\rm \Delta \log(t)=-0.403$, $\rm \Delta \log(Z/Z_{\odot})=-0.130$, and
$\rm \Delta [Mg/Fe]=-0.280$; for NGC~4649 $\rm \Delta \log(t)=0.198$, $\rm
\Delta \log(Z/Z_{\odot})=0.135$, and $\rm \Delta [Mg/Fe]=0.124$. The stellar
content of M32 is less metal rich, less enhanced in $\alpha$-elements, and much
younger (or more safely has a much younger component contributing to $\rm
H_{\beta}$) than NGC~4649.

\begin{figure}
\psfig{file=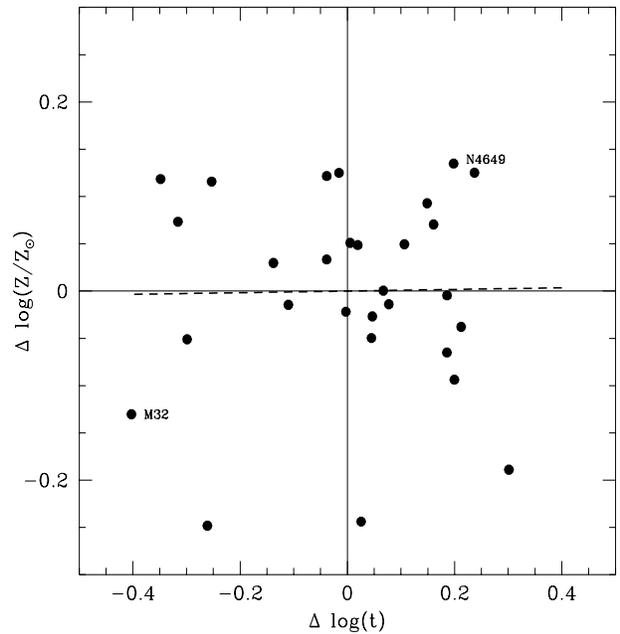,height=9.0truecm,width=8.5truecm}
\caption{The $\rm \Delta \log(Z/Z_{\odot})$ versus $\rm \Delta \log(t)$
relation. The {\em thick dashed line} is the linear regression of the results.
The position of M32 and NGC~4649 are indicated.}
\label{log_t_z}
\end{figure}

\begin{figure}
\psfig{file=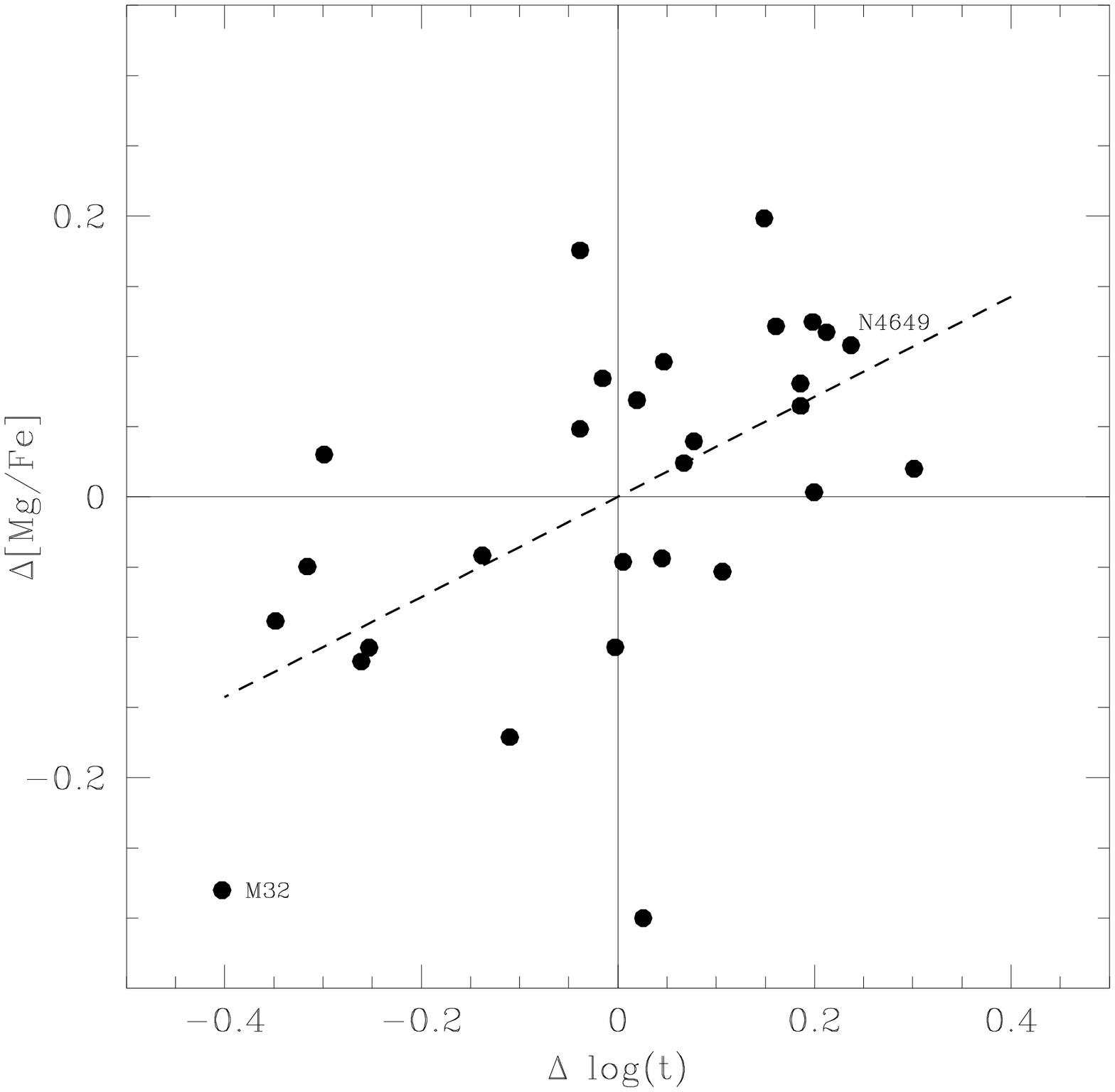,height=9.0truecm,width=8.5truecm}
\caption{The $\rm \Delta [Mg/Fe])$ versus $\rm \Delta \log(t)$ relation. The
{\em thick dashed line} is the linear regression of the results. The position
of M32 and NGC~4649 are indicated.}
\label{log_t_mgfe}
\end{figure}

\begin{figure}
\psfig{file=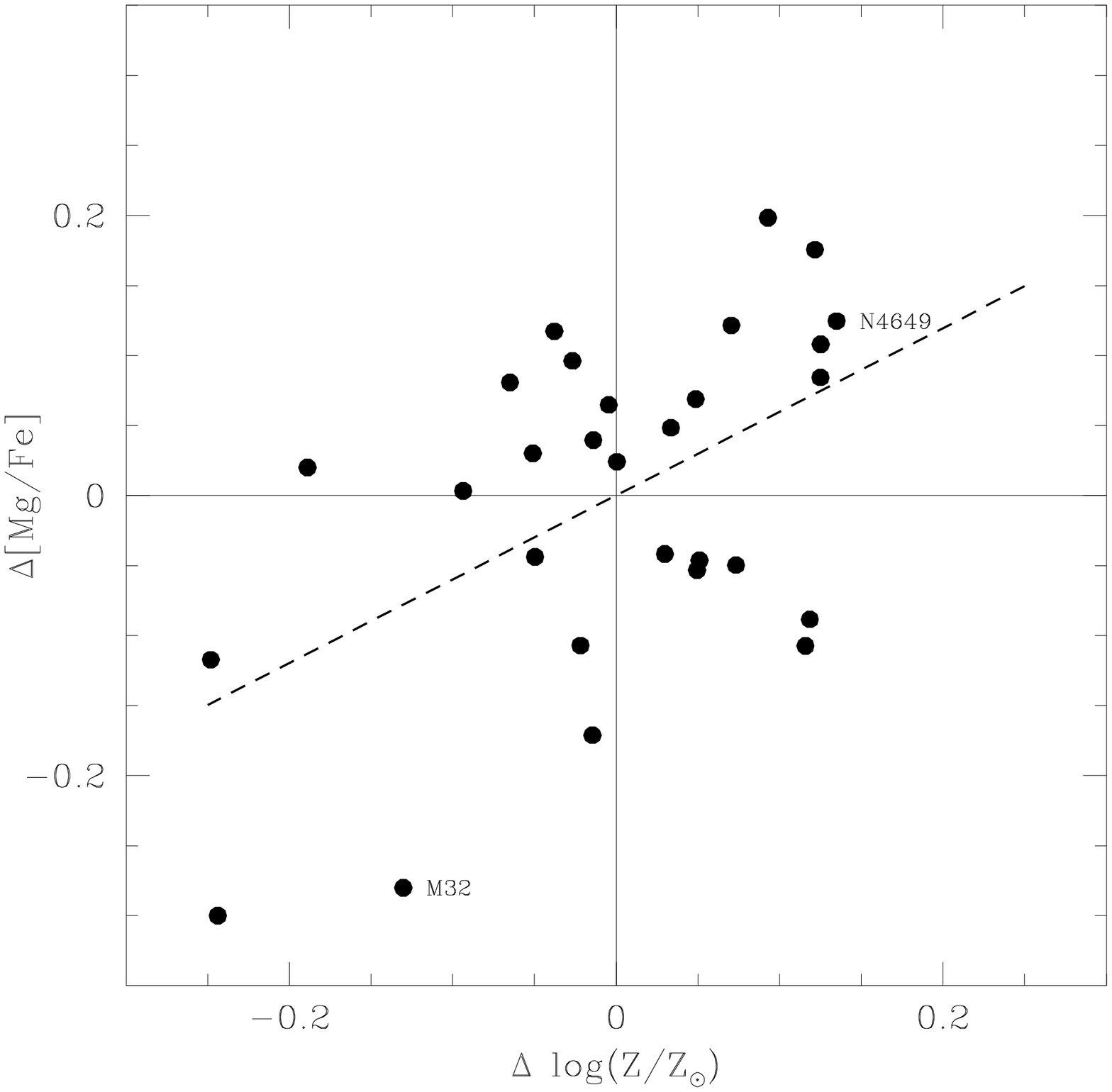,height=9.0truecm,width=8.5truecm}
\caption{The $\rm \Delta [Mg/Fe]$ versus $\rm \Delta \log(Z/Z_{\odot})$
relation. The {\em thick dashed line} is the linear regression of the results.
The position of M32 and NGC~4649 are indicated.}
\label{log_z_mgfe}
\end{figure}

\subsection{ The $\rm \Delta \log(t)$, $\rm \Delta \log(Z/Z_{\odot})$, 
             $\rm \Delta [Mg/Fe]$ space}

Aim of this section is to investigate whether systematic correlations exist
among the three physical quantities  $\rm \Delta \log(t)$, $\rm \Delta
\log(Z/Z_{\odot})$, and $\rm \Delta [Mg/Fe]$. The occurrence of such relations
would bear very much on the past history of star formation and chemical
enrichment, and the mechanism of galaxy formation as well.

Before starting this part of the analysis it is worth clarifying the real
meaning of the {\it age } parameter  $\rm \Delta \log(t)$. This age (being
mainly controlled by $H_{\beta}$) actually refers to the younger components of
the stellar mix in the galaxy (nucleus). In the ideal case (however possible)
of continuous star formation from the early epochs it would measure the
overall duration of the star forming activity. In the case that recurrent
episodes have taken place (such as in mergers and/or bursts), it would measure
the age of the last episode. Given these premises, we examine the following
relationships:

\littleskip
{\it Age-Z}. The relation between  $\rm \Delta \log(Z/Z_{\odot})$ and $\rm
\Delta \log(t)$ is shown in Fig.~\ref{log_t_z}, where only a large scatter is
seen (no systematic trend). The degree of metal enrichment seems to be totally
unrelated to the age, in the sense that at any given metallicity all ages are
possible. Does it means that sporadic episodes of star formation may change
the age parameter without significantly affecting the metallicity? It is worth
recalling that $\rm H_{\beta}$ is very sensitive to recent star formation, and
that a burst of stellar activity implying even a very small fraction of the
galaxy mass would immediately increase $\rm H_{\beta}$ without affecting the
metallicity (cf. Bressan et al. 1996). Furthermore, the recovery time of $\rm
H_{\beta}$ after a burst is of order of about 1 Gyr, after which  no trace of
the star forming period would be easily detectable with the $\rm H_{\beta}$
diagnostic.
\littleskip

{\it Age-[Mg/Fe]}. The relationship between $\rm \Delta [Mg/Fe]$ and $\rm
\Delta \log(t)$ is shown in Fig.~\ref{log_t_mgfe}. Now a good relation is
found: older galaxies seem to be more enhanced in $\alpha$-elements.
\littleskip

{\it Z-[Mg/Fe]}. The relationship between $\rm \Delta [Mg/Fe]$ and $\rm \Delta
\log(Z/Z_{\odot})$ is shown in Fig.~\ref{log_z_mgfe}. It appears that more
metal-rich  galaxies are also more enhanced in $\alpha$-elements.
\littleskip

\begin{figure}
\psfig{file=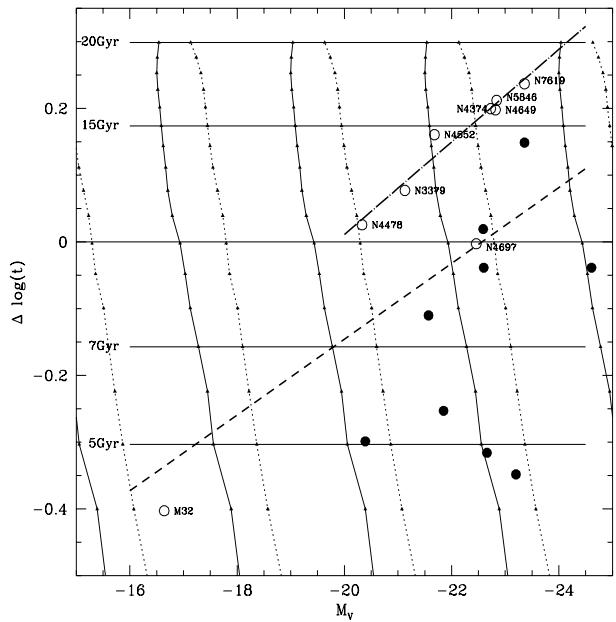,height=9.0truecm,width=8.5truecm}
\caption{The $\rm \Delta \log(t)$ versus $\rm M_V$ relation. The {\em thick
dashed line} is the linear regression of all the results. The {\em long-dashed
dotted line} is the linear regression of the the data relative to the galaxies
indicated by open circles (see the text for details). The position of M32,
NGC~4649 and other {\it quiescent} galaxies (four of which are  in the Virgo
cluster) are indicated by the open symbols. Superposed to this diagram is are
the fading lines of SSPs: the thin dotted and solid lines are for Z=0.004 and
Z=0.05, respectively; each dot along the lines show the age in step of 1 Gyr
starting from 20 Gyr (top) down to 4 Gyr (bottom); the thin horizontal lines
locate the loci of constant age (20, 15, 10, 7, and 5 Gyr starting from the
top). Finally, the fading lines of SSP (in pairs because of the different
metallicity) are shown for different values of the total mass, namely
$10^{12}$, $10^{11}$, $10^{10}$, $10^9$, and $10^8\rm \times M_{\odot}$ from
right to left.}
\label{age_mv}
\end{figure}

\begin{figure}
\psfig{file=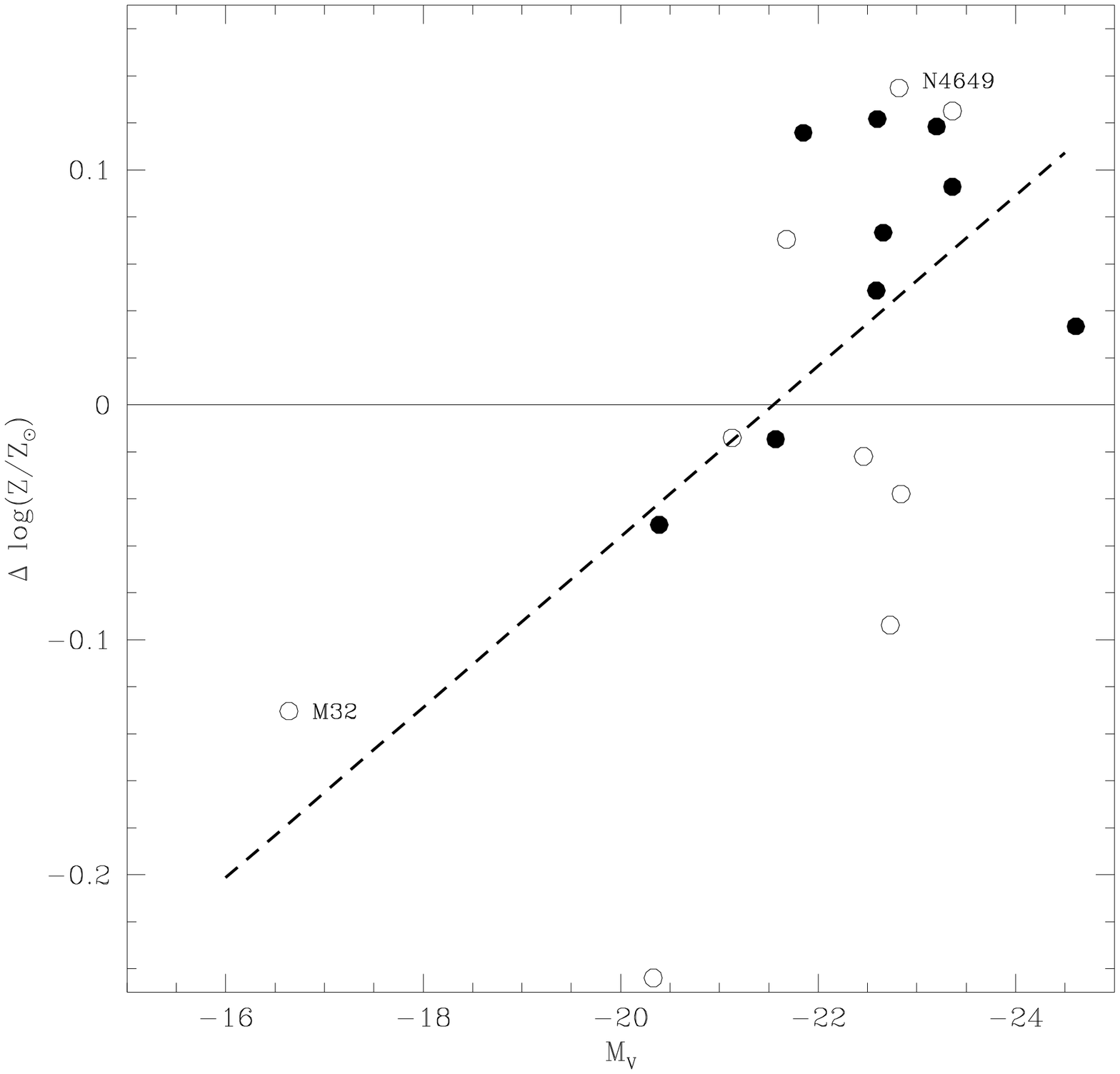,height=9.0truecm,width=8.5truecm}
\caption{The $\rm \Delta \log(Z/Z_{\odot})$ versus $\rm M_V$ relation. The
{\em thick dashed line} is the linear regression of the results. The position
of M32, NGC~4649 and other {\it quiescent} galaxies (four of which are  in the
Virgo cluster) are indicated by the open symbols.}
\label{zeta_mv}
\end{figure}

\begin{figure}
\psfig{file=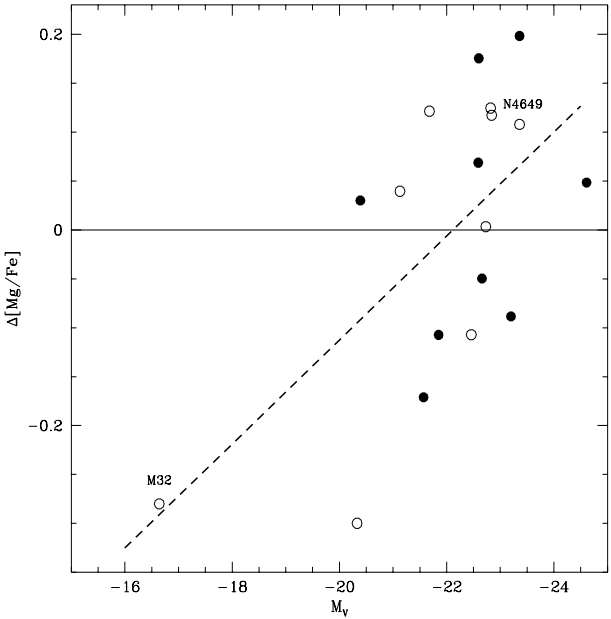,height=9.0truecm,width=8.5truecm}
\caption{The $\rm \Delta [Mg/Fe]$ versus $\rm M_V$ relation. The {\em thick
dashed line} is the linear regression of the results. The position of M32,
NGC~4649 and other {\it quiescent} galaxies (four of which are  in the Virgo
cluster) are indicated by the open symbols.}
\label{mgfe_mv}
\end{figure}

In order to cast light on the physical implications of the above relations, we
correlate each of three quantities $\rm \Delta \log(t)$, $\rm \Delta
\log(Z/Z_{\odot}$), and $\rm \Delta [Mg/Fe]$ to the total luminosity of the
galaxy as measured by the absolute magnitude $M_V$. The discussion below would
not change using $\rm M_B$. The three relations are shown in Fig.~\ref{age_mv}
(age), ~\ref{zeta_mv} (metallicity), and ~\ref{mgfe_mv} ([Mg/Fe]). 
\littleskip

{\it Age-$M_V$}. Inspecting the distribution of galaxies in Fig.~\ref{age_mv}
we notice that in spite of the large scatter a important trend can be seen.
First  in our sample all galaxies but  M32 are brighter than $\rm M_V=-20$,
which is the range of the classical color-magnitude relation of Bower et al.
(1992). Second among the galaxies defining the upper edge of the distribution
in age three of them, namely NGC~4374, NGC~4478, and NGC~4552, belong to the
Virgo cluster, and according to Sweitzer \& Seitzer (1992) they have either no
sign or some evidence of mild interaction or rejuvenation. In the same region
there are four more objects: NGC~3379 and  NGC~7619 which have no sign of
interaction (Sweitzer \& Seitzer 1992), and NGC~4649 and NGC~5846, for which
no special information is available. Finally, the last Virgo galaxy in our
sample, namely NGC~4697, deviates from the relation defined by its companions
in spite of its similar $\rm M_V$ and no signs of interaction (Sweitzer and
Seitzer 1992). But for this latter case, all the above galaxies seem to
cluster on a fairly tight relationship in which  the last episode of star
formation occurred earlier and earlier at increasing luminosity (likely mass
of the galaxy). Is this the locus of non-interacting galaxies, along which we
see the pristine star formation? If so, recalling the meaning of the age
parameter in usage here this implies that star formation either started later
or continued longer at decreasing galaxy mass (cf. Bressan et al. 1996 for a
similar suggestion).

All other galaxies are much scattered along the age axis. Indeed most of them
have sign of interaction or rejuvenation (Sweitzer \& Seitzer 1992).  Does it
imply that more recent star formation has occurred altering $\rm H_{\beta}$
and age assignment in turn ?

M32 is an ambiguous case because either it could represent the continuation of
the trend shown by the old galaxies, in which star formation lasted till a
recent past or it has been rejuvenated by a recent episode. As compared to
NGC~4649 there is factor 4.0 in between. Assuming the canonical age of 15 Gyr
for the oldest galaxies, M32 terminated its star formation history or suffered
from star formation about 3.75 Gyr ago.

In both cases the relatively young age of at least part of its stellar content
is also indicated by  studies of the classical color-magnitude diagram of the
resolved stars (cf. Grillmair et al. 1996 and references therein). Grillmair
et al. (1996) find that the CMD of M32 is consistent with a luminosity
weighted age of about 8.5 Gyr and [Fe/H]=-0.25. There is however a significant
component with [Fe/H]=0 for which younger ages cannot be excluded. Indeed an
age of about 6 Gyr was suggested by O'Connell (1988 and references), of 5 Gyr
by Freedman (1989), $4\div 5$ Gyr by Freedman (1992). See also Elston \& Silva
(1992), Davidge \& Jones (1992), and Hardy et al. (1994). Bressan et al.
(1994) addressed the question of the M32 age using the spectral synthesis
technique and reached the important conclusions: the bulk of stars have ages
as old as those typical of globular clusters say in the range $13\div 15$ Gyr,
consistently with Grillmair's et al. (1996) estimate; there is a younger
component which cannot be older than 5 Gyr and younger than 1 Gyr. This latter
boundary is set by the UV properties of M32 which has (1550-V)=4.5 (Burstein
et al. 1988). See Bressan et al. (1994) for all other details.
\littleskip

{\it Z-$M_V$}. The relationship between $\rm \Delta \log(Z/Z_{\odot})$ and $\rm
M_V$ is shown in Fig.~\ref{zeta_mv}: the metallicity seems to increase with
the luminosity (mass) of the galaxy. Limiting the inspection to the group of
galaxies that where used to argue about the old age limit in Fig.~\ref{age_mv},
namely NGC~3379, NGC~4374, NGC~4478, NGC~4552, NGC~4649, NGC~5846, and
NGC~7619, they seem to cluster on the mean value but for NGC~4478, which is even
more metal-poor than M32. In any case, also for this subset of galaxies the
above trend is present. Remarkable is the case of all galaxies brighter than
$\rm M_V=-20$ and apparently younger than the mean age (cf. Fig.~\ref{age_mv})
whose metallicity is in contrast significantly above the mean value. Does this
suggests that galaxies suffering from subsequent episodes of star formation
further increase their metallicity ? Finally, noticeable is the case of
NGC~315, the brightest galaxy in the sample, which has age and metallicity only
slightly below and above the mean, respectively. This somewhat weakens the
notion that bright galaxies are also metal rich.
\littleskip

{\it [Mg/Fe]-$M_V$}. Fig.~\ref{mgfe_mv} shows the relation between $\rm \Delta
[Mg/Fe]$ and $\rm M_V$. We start noticing that with the exception of NGC~4478,
all other galaxies of the group defining the old age limit in
Fig.~\ref{age_mv} are enhanced in $\alpha$-elements, whereas the remaining
galaxies have a different degree of enhancement. It appears that not
necessarily galaxies with high metallicity are also enhanced in
$\alpha$-elements, even if from the results shown in Fig.~\ref{log_z_mgfe}
some trend of this kind holds on a very broad sense. This can be  explained 
recalling that star formation over periods of time longer than about 1 Gyr
easily wipes out the signature type II supernovae in the abundance ratio $\rm
[\alpha/Fe]$ (cf. our model galaxies discussed in section 2).

\section{Age ranking}

Having established the relative differences $\rm \Delta \log(t)$, $\rm \Delta
\log(Z)$, and $\rm \Delta [Mg/Fe]$ from galaxy to galaxy, the absolute ranking
of the central regions of different galaxies can be attempted. The zero point
of the ranking scale is M32 for which we have independent estimates of the
age, metallicity, and perhaps [Mg/Fe]. We adopt the age of $3.5\div 4$ Gyr,
[Fe/H]=-0.25, and [Mg/Fe]=0.  The assignment of absolute ages is made with the
aid of grids of  SSP as drawn in Fig.~\ref{age_mv}: the thin dotted and solid
lines are for Z=0.004 and Z=0.05, respectively; each dot along the lines show
the age in step of 1 Gyr  starting from 20 Gyr (top) down to 4 Gyr (bottom);
the thin horizontal lines locate the loci of constant age (20, 15, 10, 7, and
5 Gyr starting from the top). Finally, the fading lines of SSP (in pairs
because of the different metallicity) are shown for different values of the
total mass, namely $10^{12}$,  $10^{11}$,  $10^{10}$, $10^9$, and $10^8 \times
M_{\odot}$  from right to left. The results are summarized in Table~7
limited to the group of galaxies that in Fig.~\ref{age_mv} define the old age
boundary. Incidentally, we note that in the case of M32 there is no
contradiction between the median age of 8.5 Gyr estimated by Grillmair et al.
(1996) looking at the distribution of stars in the CMD and the significantly
younger age of $3.5\div 4$ Gyr that we have adopted.  This in fact corresponds
to the youngest generation of stars in M32 whose traces are perhaps the
brightest  AGB stars  in the CMD. If this young component is really present it
would dominate the contribution to $\rm H_{\beta}$, whereas the old one would
scarcely contribute to it. With this assumption for the zero point of the age
scale,  the oldest galaxies in the sample have the plausible age of $16\div
17$ Gyr. In contrast,  had we assumed the age of 8.5 Gyr as the zero point of
our age scale, the age of the oldest galaxies would result unacceptably large
(above 30 Gyr).

\begin{table}
\begin{center}
\caption{Estimated ages, metallicities, and enhancements of $\alpha$-elements
         for the group of galaxies defining the old age edge in Fig.~17 }
\begin{tabular}{l| c r r  }
\hline
\hline
 & & & \\
\multicolumn{1}{l|}{NGC} &
\multicolumn{1}{c}{Age} &
\multicolumn{1}{c}{$<Z>$} &
\multicolumn{1}{c}{[Mg/Fe] }\\ 
 & & & \\
\hline
 & & & \\
221   &   $3.5\div 4.0$  &  0.0128 &  0.00   \\
3379  &   $11\div 12$    &  0.0167 &  0.22   \\
4374  &   $15\div 16$    &  0.0139 &  0.19   \\
4478  &   $10\div 11$    &  0.0098 & -0.01   \\
4552  &   $14\div 15$    &  0.0202 &  0.40   \\
4649  &   $15\div 16$    &  0.0235 &  0.41   \\
4697  &   $ 9\div 10$    &  0.0164 &  0.18   \\
5846  &   $13\div 14$    &  0.0158 &  0.40   \\
7619  &   $16\div 17$    &  0.0230 &  0.39   \\
 & & & \\
\hline
\hline
\end{tabular}
\end{center}
\label{tab7}
\end{table}

\section { Isolation or mergers? }

The fading lines in Fig.~\ref{age_mv} help to visualize in the age-magnitude
diagram the path followed by an evolving galaxy under a number of different 
scenarios. Perhaps the most intriguing question to address about galaxies is
which of the two main avenues for their formation and evolution has prevailed
as the dominant mechanism, i.e. isolation or mergers.

If a galaxy forms and evolves in isolation, and suffers from a number of
episodes of star formation (from one to several or even continuous) it would
simply slide up and down along its fading line  according to the stage of
stellar activity at which the galaxy is detected with little transversal shift
caused by chemical enrichment. Complicacies due to the possible presence of gas
can be neglected here. As already said the data are compatible with
this scheme suggesting that there are two groups of galaxies: those with no
sign of rejuvenation in which either the overall duration or the epoch of the
last episode of star formation seems to go inversely proportional to the
galaxy mass, and those with signs of rejuvenation which are obviously
scattered in this diagram.

The case of hierarchical merging is more difficult to discuss because the path
is determined by the mass and evolutionary stage of the merging galaxies, and
the amount of star formation taking place during  the merger event. Let us
suppose for the sake of simplicity that two identical units (same age, same
composition, and say $10^{10} \times M_{\odot}$ mass each) merge  triggering
star formation in the composite system. The resulting galaxy will brighten by
0.75 magnitude because of the increased mass, shift  to the  line for the new
total mass, slide down along this because of star formation, and slightly
redden because of the increased metallicity (passing from Z=0.001 to Z=0.05
$\rm \Delta M_V = 1.0$ magnitudes at most). The total displacement vector
depends on the relative amount of gas turned into stars. In any case, the
displacement will be nearly vertical along the $2\times 10^{10} M_{\odot}$
line in our example. As soon as star formation is over, the galaxy will fade 
at slowing rate back to a point near to the original position $(M_V - 0.75)$.
It is worth recalling that the recovery time after a burst of stellar activity
is short (going from a few $10^8$ yr  to 1 Gyr at most. In principle there is
no contradiction between this scenario and the bunch of galaxies scattered in
this diagram toward much younger ages. The problem is with they recovering
position if this latter corresponds to the group of galaxies with no sign of
rejuvenation. Indeed if the seed galaxies had the same age we would expected
the daughter galaxies to cluster along a nearly horizontal line in this
diagram. In contrast, it seems as if bright, high mass galaxies had their
merger adventure long ago, whereas the faint, low mass objects did it in a
more recent past, but starting from seed galaxies with small stellar content.
This is equivalently to say that the bulk of star forming activity took place
more and more toward the present at decreasing mass of the galaxy.
Furthermore, {\it why do not we see in this diagram any low mass galaxy (the
potential seeds of a bigger galaxy in the hierarchical scheme) at the same age
range of the big galaxies ? Is this only due to selection effects in our
sample (because no dwarf galaxies but M32 are present)  or  more subtle causes
are to be considered? } As nowadays answering this question is difficult owing
to the lack of sufficient data with precise measurements of $\rm H_{\beta}$,
$\rm Mg_2$ and $\rm \langle Fe \rangle$.

\section{ SN-driven winds or variable IMF? }

To explain the color-magnitude relation (CMR) of early type galaxies (cf.
Bower et al. 1992 for the case of the Virgo and Coma galaxies), long ago Larson
(1974) postulated that  it is the consequence of SN-driven galactic winds. In
the classical scenario, isolation and constant IMF, massive galaxies eject
their gaseous content much later and get higher mean metallicities than the
low mass ones. The implication is that the global duration of the star forming
activity is proportional to the galaxy mass, contrary to what required by the
simplest interpretation of the $\alpha$-enhancement problem (cf. Matteucci
1997 for a recent review of the subject). The present analysis seems to
confirm that on the average more luminous galaxies are more enhanced in
$\alpha$-elements. The trend  holds for those with no sign of rejuvenation
(cf. Fig.~\ref{mgfe_mv}) at least as it stems from comparing NGC~4478 to the
remaining galaxies in this group. The situation is less clear for the others.
However, taken as face values, the results of our analysis weaken the
classical SN-driven wind model. Furthermore,  the CMR requires  that more
massive galaxies are indeed more metal-rich than the low mass ones.  The
alternative that the CMR can be understood as an age sequence in the sense
that blue galaxies have started star formation much later than in red ones has
be proved to disagree with the redshift evolution of early type galaxies in
the HST Deep Field (cf. Kodama \& Arimoto 1996). As examined in the previous
section the merger scenario has some intrinsic difficulties with the
distribution of galaxies in the age-luminosity plane, and perhaps the
predictions for the narrow band indices and chemical properties (cf. the
experiment made with Model-C).

It follows from all this that the only viable solution of the puzzle is a 
scheme in which all galaxies have begun to form stars at the same time but,
depending on their mass, the process has continued (maybe in discrete
episodes) over different periods of time. Massive galaxies did it in the far
past, the star forming period ceased very soon, the material has been
significantly enriched in metals and $\alpha$-elements. Low mass galaxies
started at the same epoch, but continued for longer periods of time (the
duration increasing at decreasing galactic mass), got less metal enriched and
$\alpha$-enhanced (for global periods of star formation longer than about 1
Gyr this is less of  a problem).

Can such a scheme result from standard assumptions concerning star formation
and IMF? The answer is not. In a recent paper Chiosi et al. (1997) have
proposed new models of elliptical galaxies in which the IMF instead of being
constant is a function of density, temperature and velocity dispersion of the
medium in which stars are formed (cf. Padoan et al. 1997). In brief,  in a
hot, rarefied medium the new IMF is more skewed towards the high mass end than
in a cool, dense medium. This kind of situation is  met passing from a high
mass (low mean density) to  a low mass (high mean density) galaxy or from the
center to the periphery of a given galaxy. The above dependence of the IMF
yield galactic models having the desired behaviour, thus providing a plausible
way out to the above points of contradiction encountered with the standard
scheme. These models in fact predict the onset of galactic winds and consequent
termination of the star forming period much earlier in massive galaxies than
in the low mass ones. The reason of it resides in the skewness of the IMF
toward the high mass end that changes  with the mean gas density (galactic
mass and/or position within a galaxy) thus favoring in massive galaxies the
relative percentage of SN explosions and consequent heating of the gas to the
escape velocity. In this scheme massive galaxies despite their short duration
of stellar activity yet reach high metallicities and  [Mg/Fe] ratios. The
opposite occurs in the low mass ones. See   Chiosi et al. (1997) for all
details. Finally, we notice that the results of the variable IMF scheme are
also compatible with the indications arising from the present study, in
particular the age ranking inferred from $\rm H_{\beta}$.

\section{Conclusions}

In this study we have thoroughly investigated the ability of the $\rm
H_{\beta}$, $\rm Mg_2$ and $\rm \langle Fe \rangle$ diagnostic to assess 
chemical abundances and their ratios and ages of elliptical galaxies. In
particular we have addressed the question whether different slopes of the
gradients in  $\rm Mg_2$ and $\rm \langle Fe \rangle$ across galaxies do
automatically imply an enhancement of the [Mg/Fe] ratio. Second, we have
tackled the problem of the real information hidden in the different values of
$\rm H_{\beta}$, $\rm Mg_2$ and $\rm \langle Fe \rangle$ observed in the
nuclear regions of elliptical galaxies. The results of this study can be
summarized as follows:

\begin{enumerate}
\item{The Line strength indices $\rm Mg_2$ and $\rm \langle Fe \rangle$ do not
simply correlate with the chemical abundances.  To infer from line strength
indices the corresponding chemical abundances (and their ratios) one needs to
know: (i) the star formation history of the galaxy; (ii) the metallicity
partition function $\rm N(Z)$; (iii) and via a suitable calibration the
effects of the  enhancement in $\alpha$-elements on the line strength indices.
This conclusion does not depend on the particular models and  calibration used
to perform our experiments, but stems from the intrinsic properties of the
line strength indices themselves.}

\item{ We provide basic calibrations for the variations $\rm \delta
H_{\beta}$, $\rm  \delta Mg_2$ and $\rm \delta \langle Fe \rangle$ as a
function of age $\rm \Delta \log(t)$ (in Gyr), metallicity $\rm \Delta
\log(Z/Z_{\odot})$,  and $\rm \Delta [Mg/Fe]$  of SSPs whose application is of
general use.}

\item{Limited to three galaxies of the Carollo \& Danziger (1994a,b) catalog we
analyze the implications of the gradients in $\rm Mg_2$ and $\rm \langle Fe
\rangle$ observed across these systems. It is shown from a qualitative point
of view how the difference $\rm \delta Mg_2$ and $\rm \delta \langle Fe
\rangle$ between the values of each index at any radial distance with respect
to their  central values would translate into the $\rm \Delta [Mg/Fe]$,  $\rm
\Delta \log(Z/Z_{\odot})$, and $\rm  \Delta \log(t)$ differences between the
local and the central values.}

\item{ The above calibration is used to explore the variation from galaxy to
galaxy of the nuclear values of $\rm H_{\beta}$, $\rm Mg_2$, and $\rm \langle
Fe \rangle$ limited to a sub-sample of the Gonzales (1993) catalog. The
differences $\rm \delta H_{\beta}$, $\rm \delta Mg_2$, and $\rm \delta \langle
Fe \rangle$ are converted into the differences $\rm \Delta \log(t)$, $\rm
\Delta \log(Z/Z_{\odot})$, and $\rm \Delta [Mg/Fe]$. Various correlations among
the age, metallicity, and enhancement variations are explored. In particular
we thoroughly examine the relationships $\rm \Delta \log(t)-M_V$, $\rm \Delta
\log(Z/Z_{\odot})-M_V$, and $\rm \Delta [Mg/Fe]-M_V$, and advance the
suggestion that the duration of the star formation period gets longer or the
last episode of stellar activity  gets closer and closer toward the present at
decreasing galaxy mass. This result is discussed at the light of predictions
from the merger and isolation models of galaxy formation and evolution. In
brief, we conclude that none of these can explain the results of our analysis,
and suggest that the kind of time and space dependent IMF proposed by Padoan
et al. (1997) and the associated models of elliptical galaxies elaborated by
Chiosi et al. (1997) should be at work.}
\end{enumerate}
 
\vskip 0.5truecm

This study has been financed by the Italian Ministry of University, Scientific
Research and Technology (MURST), the Italian Space Agency (ASI), and the TMR
grant ERBFMRX-CT96-0086 from the European Community.


\newpage
\newpage

\end{document}